\documentclass[a4paper,11pt]{article}
\pdfoutput=1 
\usepackage{jheppub} 
\usepackage[T1]{fontenc} 
\usepackage{amsmath}
\usepackage{amssymb}
\usepackage{stmaryrd} 
\usepackage{amsfonts}
\usepackage{graphicx}
\usepackage{enumitem}
\usepackage{bm}
\usepackage{rotating}
\usepackage{hyperref}
\usepackage{pbox}
\usepackage[dvipsnames]{xcolor}
\usepackage{array}
\usepackage{physics}
\usepackage{bm}
\usepackage{dsfont}
\usepackage{mathtools}
\usepackage{leftidx}
\usepackage{lmodern}
\usepackage{braket}
\usepackage{tensor}
\usepackage{mathrsfs}

\usepackage{float}
\usepackage[title]{appendix}
\usepackage[normalem]{ulem}
\usepackage{mathtools}
\usepackage{tikz}
\usetikzlibrary{arrows.meta,decorations.pathmorphing,math}

\newcommand{\tf}{\tilde{f}}
\newcommand{\tN}{\tilde{N}}
\newcommand{\tnu}{\tilde{\nu}}
\newcommand{\tgamma}{\tilde{\gamma}}

\numberwithin{equation}{section}

\newcommand{\be}{\begin{equation}}
\newcommand{\ee}{\end{equation}}
\newcommand{\ba}{\begin{eqnarray}}
\newcommand{\ea}{\end{eqnarray}}

\newcommand{\beq}{\begin{equation}}
\newcommand{\eeq}{\end{equation}}
\newcommand{\beqa}{\begin{eqnarray}}
\newcommand{\eeqa}{\end{eqnarray}}
\newcommand{\nn}{\nonumber}

\newcommand{\llb}{\llbracket }
\newcommand{\rrb}{\rrbracket }
\newcommand{\SN}{{\mbox{\tiny SN}}}

\title{On a lower-dimensional Killing vector origin of irreducible Killing tensors
}

\date{today}

\author[a]{Finnian Gray,}
\emailAdd{finnian.gray@univie.ac.at}

\author[b]{Gloria Odak}
\emailAdd{gloria.odak@matfyz.cuni.cz}

\author[b]{Pavel Krtou{\v s}}
\emailAdd{pavel.Krtous@utf.mff.cuni.cz}

\author[b]{David Kubiz\v n\'ak,}
\emailAdd{david.kubiznak@matfyz.cuni.cz}

\affiliation[a]{Faculty of Physics, University of Vienna, Boltzmanngasse 5, 1090, Vienna, Austria}

\affiliation[b]{Institute of Theoretical Physics, Faculty of Mathematics and Physics,
Charles University, Prague, V Hole{\v s}ovi{\v c}k{\' a}ch 2, 180 00 Prague 8, Czech Republic}

\abstract{
Considering a spacetime foliated by co-dimension-2 hypersurfaces, we find the conditions under which lower-dimensional symmetries of a base space can be lifted up to irreducible Killing tensors of the full spacetime. 
In this construction, the key ingredient for irreducibility is the non-commutativity of the underlying Killing vectors. 
It gives rise to a tower of growing rank Killing tensors determined by the structure constants of the corresponding Lie algebra. 
A canonical example of a metric with such emergent non-trivial hidden symmetries in all dimensions is provided by rotating (off-shell) generalized Lense--Thirring spacetimes, where the irreducible Killing tensors arise from the underlying spherical symmetry of the base space. 
A physical on-shell realization of this  construction in four dimensions is embodied by a rotating black hole in the Einstein--Maxwell-Dilaton-Axion theory. 
Further examples of equal spinning Myers--Perry spacetimes and spacetimes built on planar and Taub-NUT base metrics are also discussed. 
}

\makeatletter
\gdef\@fpheader{}
\makeatother

\begin{document} 

\maketitle
\flushbottom

\section{Introduction}

Symmetries play an important role in the study of dynamical systems. 
In general relativity, {\em explicit} spacetime symmetries are characterized by Killing vector fields, which generate conserved quantities and simplify the analysis of geodesic motion and field equations. However, beyond these explicit symmetries, certain spacetimes exhibit \emph{hidden symmetries}, encoded in higher-rank tensor fields known as {(symmetric) {\em Killing tensors}~\cite{Walker:1970un} and (antisymmetric)} {\em Killing--Yano tensors}~\cite{yano1952some}. 
Such symmetries are known to underlie remarkable properties of rotating black holes described by Kerr geometry~\cite{kerr1963gravitational} and its higher-dimensional generalizations~\cite{Myers:1986un,Gibbons:2004js,Gibbons:2004uw}, see e.g. \cite{Carter:1968rr,Carter:1977pq,Frolov:2006dqt,Kubiznak:2006kt,Frolov:2017kze} for details and references. 
{In this work we focus on hidden symmetries encoded in Killing tensors.}

{A Killing tensor of rank-$p$ is a completely symmetric tensor, $K^{\mu_1\dots \mu_p}=K^{(\mu_1\dots \mu_p)}$, obeying the following Killing tensor equation: 
\be\label{KT} 
\nabla^{(\nu}K^{\mu_1\dots \mu_p)}=0\,.
\ee 
Incorporating Killing vectors as $p=1$ subcase, Killing tensors are in one-to-one correspondence with monomial constants of geodesic motion---obtained by contracting all indices of the Killing tensor with the corresponding geodesic momenta $p_\mu$, $p^\nu\nabla_\nu p^\mu=0$, namely
\be \label{constants}
C_p=K^{\mu_1\dots \mu_p}p_{\mu_1}\dots p_{\mu_p}\,. 
\ee
The Poisson bracket on these constants induces the so called {\em Schouten--Nijenhuis (SN) bracket} \cite{schouten1940uber,nijenhuis1955jacobi} on Killing tensors (see Appendix~\ref{App: SN brackets}). 
This gives rise to an algebra of Killing tensors (see Refs. \cite{dolan1989significance,barbance1973tenseurs,DeLong1982,Takeuchi1983,Thompson1986,MCLENAGHAN2004621,Yue2005,michel2017prolongation} for details on the algebraic structure), and in principle allows one to generate new (higher rank) Killing tensors via the SN bracket.
If, for a geodesic motion in $d$ spacetime dimensions, one can find $d$ functionally {\em independent} and mutually Poisson commuting constants of motion \eqref{constants}, the motion is {\em completely integrable}. 

To capture the `independence' of the corresponding Killing tensors, the following notions have often been used. 
Killing tensors are called {\em reducible}, if they can be decomposed as a linear combination of a symmetrized product of lower rank Killing tensors. 
Otherwise they are called {\em irreducible} \cite{Walker:1970un,dolan1989significance}. Although this terminology is not particularly useful when it comes to the integrability purposes%
\footnote{Reducible Killing tensors obviously do not imply the existence of new independent constants of geodesic motion and thence can be discarded.  
However, the above notion of irreducibility itself does not imply that the corresponding constants of geodesic motion are automatically functionally independent (as complicated non-polynomial relations may exist among such constants). 
Note also that, whereas the upper bound on independent constants of motion in $d$ dimensions is $2d-1$ (corresponding to a {\em maximally superintegrable system}, see e.g. \cite{Miller_2013}), the number of rank-$p$ Killing tensors can be larger, see \eqref{eq: Kmax} below.
}
we will keep using it in this work. 
}

Recently, in a series of papers \cite{Gray:2021toe, Gray:2021roq, Gray:2024qys}, it was argued that  `magic square' {\em Lense--Thirring spacetimes}, first introduced in ~\cite{Baines:2020unr,Baines:2021qaw} (see also \cite{Sadeghian:2022ihp}), admit a growing tower of higher-rank Killing tensors. 
These tensors, constructed via the SN bracket, form {a growing (quadratically with the number of spacetime dimensions)} sequence of Killing tensors, giving rise to a spacetime with more hidden symmetries than explicit ones. 
While these results suggest a deep underlying algebraic structure, a clear geometric and physical explanation of this phenomenon remains an open question.

In this work, we aim to demystify these results by proposing a systematic construction of irreducible Killing tensors from lower-dimensional Killing vectors. 
Similar such ideas have appeared earlier. 
We refer the reader to~\cite{Beig:1996ew,Chrusciel:2013aya} for work on conditions under which Killing initial data (i.e. Killing vectors on a hypersurface--spacelike or characteristic) extend to the full spacetime using the Einstein equations.
On the other hand, the works \cite{Kobialko:2021aqg, Bogush:2021qnz} have considered a foliation of  co-dimension-1 hypersurfaces and a geometric procedure to lift symmetries of the hypersurface to the full spacetime. 
{Similarly, in \cite{Kubiznak:2007ca} the Killing tensors of the off shell Kerr--NUT-AdS class of spacetimes have been shown to be Killing tensors of the induced metric on the $t=$const. co-dimension-1 foliation.}
See also \cite{Garfinkle:2010er,Garfinkle:2013cha} for similar ideas (the {latter} with an application to Killing--Yano tensors), {\cite{rajaratnam2014killing,Krtous:2015ona} for studies of the lift of hidden symmetries on warped spacetimes, and \cite{Gibbons:2011hg} for the Eisenhart lift of lower-dimensional hidden symmetries to higher-rank Killing tensors.}

In contrast, we consider a spacetime in which an off-shell geometry is foliated by \emph{codimension-2 hypersurfaces} and demonstrate that the {\em non-commutativity} of Killing vectors in the lower-dimensional base space comes into play when generating a hierarchy of Killing tensors. 
The structure constants of the underlying Lie algebra dictate the emergence of a growing tower of Killing tensors in the full spacetime. 
To illustrate this general construction, we explore its realization in the aforementioned generalized Lense--Thirring spacetimes.
Furthermore, we identify an explicit on-shell realization of our framework in the context of {\em Einstein--Maxwell-Dilaton-Axion (EMDA)} black hole {solutions  \cite{Clement:2002mb}.} 
We also show that the well-known $U([d-1]/2)$ symmetry of the Myers-Perry black holes in $d$ dimensions with equal rotation parameters~\cite{Vasudevan:2004ca} can be understood as arising from the Killing vector symmetries of the base space.
Finally, to {further} illustrate the construction we consider two {(mathematical)}  examples: a lifting of a planar metric and the Euclidean Taub--NUT instanton to a Lorentzian spacetime in two dimensions higher.

The structure of this paper is as follows. 
In Section~\ref{sec: Generic Metrics}, we introduce {our metric ansatz} and outline the formalism for lifting Killing vectors to Killing tensors. 
In Section~\ref{sec: LT}, we apply our framework to generalized Lense--Thirring spacetimes---demonstrating the construction of Killing tensors {in arbitrary number of spacetime dimensions and presenting an explicit on-shell realization in the EMDA theory in 4D. Further examples are discussed in Sec.~\ref{Sec4}.
We conclude in Sec.~\ref{sec: conclusion}. 
Additional material has been moved to the appendices. 
Namely, Appendix~\ref{app: foliation} reviews the construction of the codimension-2 foliation, Appendix~\ref{app: dual met} presents an alternative construction with `dual metric ansatz', and Appendix~\ref{App: SN brackets} reviews the properties of the SN brackets.   
}

\section{Lifting symmetries from codimension-2 hypersurfaces}
\label{sec: Generic Metrics}
{Let us consider a $d$-dimensional spacetime $({\cal M}, g)$, foliated by a (Riemannian) codimension-2 hypersurface $(\cal S,\gamma)$, which we will sometimes refer to as the `base space'. Our aim is to  investigate the conditions under which the full spacetime `inherits' interesting symmetry properties from ${\cal S}$. }

\subsection{Metric ansatz}

{
In what follows, we will use the Greek alphabet $\mu,\nu,\rho,\dots$ to label coordinates $x^\mu$ and the indices of tensors in the full spacetime ${\cal M}$ and the uppercase Latin script $A,B,C,\dots$ to label coordinates $x^A$ and the indices of tensors intrinsic to codimension-2 hypersurface ${\cal S}$.}
{We further assume the spacetime metric takes the following form:}
\be\label{Generic}
g=g_{\mu\nu}dx^\mu dx^\nu=-Nfdt^2+\frac{dr^2}{f}+\gamma_{AB}(dx^A+\nu^A dt)(dx^B+\nu^B dt)\,.
\ee
{Here, $\partial_t$ and $\partial_r$ are (decoupled) timelike and radial directions,} 
$N=N(t,r)$ and $f=f(t,r)$ are arbitrary functions of $t$ and $r$, and $\gamma_{AB}=\gamma_{A B}(t,r, x^C)$ and $\nu^A=\nu^{A}(t,r,x^C)$ are the metric and  an arbitrary vector on $\cal S$.%
\footnote{\label{FN: foliation}
While $t,r$ are coordinates on the full spacetime, they are `external parameters' for objects on ${\cal S}$ such as $\gamma_{AB}$ and $\nu^{A}$. 
In fact, this means that we truly consider a family of hypersurfaces ${\cal S}^{(r,t)}$ together with the corresponding family of intrinsic metrics $\gamma^{(r,t)}$. This subtlety does not play any significant role in what follows.  
Moreover, if $f$ and $N$ are allowed to depend on $x^A$ as well then the metric \eqref{Generic} is valid (locally) for any generic spacetime---although some gauge fixing is used, see Appendix \eqref{app: foliation} for details.
}
The inverse metric is given by
\be\label{Generic inverse}
g^{-1}=g^{\mu\nu}\partial_{\mu}\partial_{\nu}=-\frac{1}{fN}(\partial_t-\nu^A\partial_{A})^2+f\partial_r^2+\gamma^{AB}\partial_A\partial_B\,,
\ee
{where $\gamma^{AB}$ is the inverse of $\gamma$. For completeness we discuss the gauge freedom of such foliations in Appendix \ref{app: foliation} and an alternative anasatz in Appendix \ref{app: dual met}.}

\subsection{Lifting the SN bracket}

{In what follows, we will employ a `{\em natural lift}' of tensors on ${\cal S}$ to ${\cal M}$. 
Namely, for a vector $\xi\in T{\cal S}$, $\xi^A=\xi^A(t,r,x^B)$, we define the corresponding `lifted' vector ${\hat \xi}\in T{\cal M}$  as:}
\be
\hat{\xi}\equiv\hat{\xi}^\mu\partial_\mu=0\times\partial_t +0\times \partial_r+\xi^A\partial_A\,.
\ee
More generally, the above lift naturally extends to any rank-$p$-tensor $X^{A_1\dots A_p}\in T^p{\cal S}$, namely:
\be \label{eq: lift}
T^p{\cal{M}}\ni \,\hat{X}\equiv X^{A_1\dots A_p}\partial_{A_1}\dots\partial_{A_p}
\,,
\ee
with any component in $t$ and $r$ directions missing. 
Due to the splitting of the metric \eqref{Generic} it can be shown that this lift preserves the inner product:
$
<X,Y>_\gamma=<\hat{X},\hat{Y}>_{g}\,.
$

{In order to lift the symmetries from ${\cal S}$ to ${\cal M}$, we shall employ the tool of SN brackets. For symmetric tensors $A^{\mu_1\dots \mu_p}$ and $B^{\nu_1\dots \nu_q}$ on $\cal M$, these are defined as follows \cite{schouten1940uber,nijenhuis1955jacobi}: 
\ba\label{SN}
[A,B]_{\mbox{\tiny SN}}^{\mu_1\dots \mu_{p+q-1}}&=&
pA^{\rho (\mu_1\dots \mu_{p-1}}\partial_\rho B^{\mu_p\dots \mu_{p+q-1})} 
 -qB^{\rho(\mu_1\dots \mu_{q-1}}\partial_\rho A^{\mu_q \dots \mu_{q+p-1})}\,.
\ea
In particular, for a vector $A=\xi$, the above reduces to the Lie derivative. 
The definition also extends naturally to scalars, $B=\Phi$ by 
\be\label{eq: SNB scalar}
[A,\Phi]^{\mu_1\dots \mu_{p-1}}_\SN=pA^{\rho \mu_1\dots \mu_{p-1}}\partial_\rho\Phi\,.
\ee
Note further, that the rank-$p$ Killing tensor equation \eqref{KT} can be conveniently expressed as 
\be
[K,g^{-1}]=0\,. 
\ee

Moreover, it follows from the definition of the SN bracket (derived from the Poisson algebra of the corresponding monomial constants of geodesic motion) that the SN bracket of two Killing tensors produces again a (possibly trivial) Killing tensor; see Appendix~\ref{App: SN brackets} for more details.  We also refer the reader to \cite{barbance1973tenseurs,DeLong1982,Takeuchi1983,Thompson1986,MCLENAGHAN2004621,Yue2005,michel2017prolongation} for the characterization of Killing tensors in spaces of constant curvature. 
}

Since we also need to distinguish brackets and symmetries on ${\cal M}$ from symmetries on ${\cal S}$,
we shall use the following notation for the SN bracket between tensors intrinsic to ${\cal S}$,
$X^{A_1\dots A_p}$ and $Y^{B_1\dots B_q}$:
\ba\label{SN2}
\llb X,Y\rrb _{\mbox{\tiny SN}}^{A_1\dots A_{p+q-1}}&=&
pX^{C (A_1\dots A_{p-1}}\partial_C Y^{A_p\dots A_{p+q-1})} 
 -qY^{C(A_1\dots A_{q-1}}\partial_CX^{A_q \dots A_{q+p-1})}\,. 
\ea
Notice that we have the natural relationship between the bracket of the lift of an object and the lift of the intrinsically defined object:
\be
[\hat{X},\hat{Y}]_{\SN}=\widehat{\llb X, Y\rrb}_{\SN}\,.
\ee

Next, a short calculation using \eqref{Generic inverse} shows we can decompose the bracket of a lifted object on ${\cal M}$ with $g^{-1}$ into the bracket on ${\cal S}$ with $\gamma^{-1}$ as follows: for a given $X^{A_1\dots A_p}$ we have 
\begin{align}\label{eq: Lifted bracket}
    [\hat{X}, g^{-1}]_{\SN}^{\mu_1\dots \mu_{p+1}}\partial_{(\mu_1\dots\mu_{p+1})}=
    &-\llb X, (fN)^{-1} \rrb^{ A_1\dots A_{p-1} }_{\SN}\,\, \partial_{(t}\partial_t\partial_{A_1}\dots\partial_{A_{p-1})}
    \nonumber\\
    &+\llb X, f\rrb^{ A_1\dots A_{p-1} }_{\SN}\,\, \partial_{(r}\partial_r\partial_{A_1}\dots\partial_{A_{p-1})} 
    \nonumber\\
    &+ 2\left((fN)^{-1}\partial_t X^{ A_1\dots A_{p}}+\llb X, (fN)^{-1}\nu \rrb^{ A_1\dots A_{p} }_{\SN} \right)\partial_{(t}\partial_{A_1}\dots\partial_{A_{p})}
    \nonumber\\
    &-2f \left(\partial_r X^{ A_1\dots A_{p}}\right) \,\, \partial_{(r}\partial_{A_1}\dots\partial_{A_{p})}
    \nonumber\\
    &-2(fN)^{-1}\nu^{(A_1}\partial_t X^{A_2\dots A_{p+1})} \partial_{(A_1}\dots\partial_{A_{p+1})}\,
    \nonumber\\
    &+\left(\llb X, \gamma^{-1} \rrb_\SN-\llb X, (fN)^{-1}\nu\otimes\nu \rrb_\SN  \right)^{A_1\dots A_{p+1}}\partial_{(A_1}\dots\partial_{A_{p+1})}\,.
\end{align}
Here, on the right hand side, $f$, $N$, and $\nu^A$ are viewed as intrinsic quantities to $\cal S$.

{Let us now assume that $X$ is a symmetry of $\gamma$, that is 
\be
 \llb X, \gamma^{-1}\rrb_{\SN}=0\,.
\ee
Then, $\hat X$} will be a symmetry of the full spacetime, i.e. $[\hat{X},g^{-1}]_{\SN}=0$, provided that i) $\llb X, f \rrb_{\SN}=0=\llb X, N \rrb_{\SN}$ (which is in our case automatically satisfied as the functions $f, N$ depend on $t$ and $r$ only), and ii)
\begin{equation}\label{eq: Symm reqs}
    \partial_t X=0\,,\quad \partial_r X=0\,,\quad \llb X, \nu \rrb_{\SN}=0\,.
\end{equation}
Notice here that requiring $X$ to be lifted to a symmetry of the full spacetime imposes no requirements for $\nu^A$ to be a symmetry of $({\cal S}, \gamma)$.%
\footnote{As mentioned in Footnote \ref{FN: foliation}, if $f$ and $N$ depend on the full set of coordinates $x^\mu=(t,r,x^A)$ the the metric \eqref{Generic} actually has enough freedom to describe (locally) any spacetime. 
Thus, \eqref{eq: Lifted bracket}, or points i) and ii), provide generic conditions for symmetries of $( {\cal S}, \gamma)$ to lift to $({\cal M},g)$ using the mapping in \eqref{eq: lift}.}

{
In particular, the above equations imply that $\hat\nu$ itself will not be a Killing vector on ${\cal M}$ as it is $(t,r)$-dependent. In what follows, we assume that $\nu$ can be decomposed as
\be\label{eq: nu ansatz}
\nu^A\equiv p(t,r)\xi_0^A(x^B)\,,
\ee
where 
$\xi_0$ is a Killing vector on ${\cal S}$, in order that our full spacetime have at least one Killing vector, namely $\hat{\xi}_0$.
}

Let $\xi_1$ be another Killing vector on $(\gamma, {\cal S})$. Then, in order that its lift ${\hat \xi}_1$ be a Killing vector of the full spacetime, the above conditions then impose that the vector $\xi_1^A$ be $(t,r)$-independent
\be
\partial_r\xi_1^A=0=\partial_t\xi_1^A\,,
\ee
and that the two Killing vectors $\xi_0$ and $\xi_1$ on ${\cal S}$ must Lie commute:%
\footnote{We will differentiate notation for the Lie bracket of a vector field on $T{\cal S}$  and $T{\cal M}$, as above. That is, $\llb\cdot, \cdot \rrb$ and $[\cdot, \cdot ]$ respectively---a bracket without a subscript SN will denote the Lie bracket (which for two vectors is the same object).}
\be\label{eq: Com KV}
\llb \xi_0,\xi_1\rrb^A=\frac{1}{p(t,r)}
\llb \nu,\xi_1\rrb^A=0\,.
\ee
{Lifting of explicit symmetries thus requires commutativity on ${\cal S}$. In what follows we shall concentrate now on the case where the lower-dimensional Killing vectors do not commute, but give rise to hidden symmetries in the full spacetime.}

\subsection{{Rank-2 irreducible Killing tensors from lower-dimensional Killing vectors}}

Suppose now that, {apart from the Killing vector $\xi_0$ \eqref{eq: nu ansatz},} we have two additional Killing vectors on $T{\cal S}$, $\xi_1^A=\xi_1^A(t,r,x^B)$ and $\xi_2^A=\xi_2^A(t,r,x^B)$, which do not necessarily commute with $\xi_0^A$.
Instead we have the Lie algebra
\be
\llb \xi_i,\xi_j\rrb=f_{ijk}\xi_k\,, \quad \text{where } i,j,k,\in\{0,1,2\}\,,
\ee
where $f_{ijk}=f_{[ij]k}$ are some 
(nonzero) structure constants. 
This implies that individually $\hat{\xi}_1$ and $\hat{\xi}_2$  are not {necessarily} Killing vectors of $g$. 
However, {suppose further}  that their `square' (which clearly commutes with $\gamma^{-1}$ due to the Leibniz rule~\ref{eq: Leibniz SN bracket}),
\be
C_{12}\equiv \xi_1\otimes\xi_1 + \xi_2\otimes\xi_2,,
\ee
also commutes with $\xi_0$ in the SN sense:
\be\label{eq: Cas Com}
\llb C_{12},\xi_0\rrb_{\SN}=0\,.
\ee

{In fact, the latter follows automatically if the} symmetry group of $\gamma$ is {\em semi-simple}. 
In that case,  the structure constants will be totally anti-symmetric 
\be 
f_{ijk}=f_{[ijk]}\,,
\ee 
provided $\{\xi_i\}$ forms a basis of the Lie algebra, {see e.g. Chapter 1 in \cite{barut1986theory},}
and $C_{12}$ is essentially the Casimir of the algebra.%
\footnote{
{In fact, in maximally symmetric spaces (constant curvature manifolds), e.g.} 
\cite{barbance1973tenseurs,DeLong1982,Takeuchi1983,Thompson1986,MCLENAGHAN2004621,Yue2005,michel2017prolongation},
a certain quotient of the universal covering algebra of the symmetry group is isomorphic to the space of Killing tensors with the Lie bracket realized by the SN bracket (see Theorem 4.13 and corollary 4.14 in  \cite{michel2017prolongation}).
}
{To show this, we use the fact that} for a symmetric {rank-$p$ tensor $X$} and $q$ vectors $\xi_j$, the SN-bracket (on $\cal S $ or $\cal M $) satisfies a Leibniz type rule (see Appendix \ref{App: SN brackets} for details)
\be\label{eq: SN sym vecs}
\Big[\overset{{q}}{\underset{\substack{j=1}}{\bigodot}}\,\xi_j,X\Big]_{\SN}
=
\sum_{i=1}^{q} [\xi_i, X]_{\SN}
\odot
\Big(\overset{{q-1}}{\underset{\substack{j=1\\ j\neq i}}{\bigodot}} \, \xi_j\Big)   \,,
\ee
where $\bigodot$ denotes the repeated symmetrized tensor product $\odot$, e.g. for two vectors  $X\odot Y=1/2(X \otimes Y+Y\otimes X)$.
Thus, employing the form of $C_{12}$ the commutation \eqref{eq: Cas Com} reads
\begin{align}\label{eq: SN brack Cas}
    \llb C_{12},\xi_0\rrb_{\SN}^{AB}
     &=2\sum_{i=1}^2 \xi_i\tensor[^{(A}]{\llb \xi_i,\xi_0\rrb}{^{B)}}=2\sum\limits_{\substack{i=1\\
     j=0}}^{2}f_{i0j} \xi_i^{(A}\xi_j^{B)} =\sum_{i,j=0}^2f_{i0j} \xi_i^{(A}\xi_j^{B)}\underbrace{=0}_{\text{if } {f_{ijk}=f_{[ijk]}}}\!\!\!\!\!\!\!,
\end{align}
as anticipated.

In any case, provided $\xi_1$ and $\xi_2$ are $(t,r)$-independent, and condition \eqref{eq: Cas Com} holds:
\be\label{conditions}
\llb C_{12},\xi_0\rrb^{AB}_{\SN}=0\,,\quad \partial_t\xi_1^A=0=\partial_t\xi_2^A\,,\quad \partial_r\xi_1^A=0=\partial_r\xi_2^A\,,
\ee
we also have $\llb C_{12},\nu \rrb_{\SN}= p(t,r)\llb C_{12},\xi_0\rrb_{\SN}=0$.
In other words, the above conditions \eqref{conditions} are sufficient to guarantee that the lift ${\hat C}_{12}$ is a Killing tensor of the full spacetime.

In summary, starting from the form of the metric \eqref{Generic} with $\nu$ given by \eqref{eq: nu ansatz} and assuming $\xi_0^A$ \eqref{eq: nu ansatz} is a Killing vector of $\gamma_{AB}$, we have shown:
\begin{enumerate}

    \item $\hat{\xi}_0$ is a Killing vector in  $({\cal M}, g)$.
    
    \item If $\xi_1$ is another $(t,r)$-independent Killing vector of $\gamma$ which commutes with $\xi_0$, i.e.
    $$\llb \xi_0,\xi_1\rrb^A=0\,,$$
    then $\hat{\xi}_1$ is a Killing vector in $({\cal M}, g)$.

    \item If instead we have two $(t,r)$-independent Killing vectors of $\gamma$ whose square SN commutes with $\xi_0$;
    $$ C_{12}=\xi_1\otimes\xi_1 + \xi_2\otimes\xi_2\,,\quad \, \llb C_{12},\xi_0\rrb^{AB}_{\SN}=0\,,$$
    then $\hat{C}_{12}$ is a Killing tensor of the full spacetime $({\cal M}, g)$.
    
\end{enumerate}
Notice that the last point extends to  any second order $C$ constructed from a finite number of Killing vectors of $\gamma$. 
Moreover, if $C$ is a rank-$p$ Killing tensor of $\gamma$ which is $(t,r)$-independent and SN commutes with $\nu$ then \eqref{eq: Lifted bracket} shows that its lift $\hat{C}$ will be a Killing tensor of $({\cal M}, g)$.

\subsection{Axisymmetric spacetimes}\label{sec: Gen ax symm}

 We would now like to adapt the above  discussion to metrics in $d$ spacetime dimensions which are symmetric about $m=[\frac{d-1}{2}]$ axes of rotation (where $[A]$ denotes the integer part of $A$).
 That is, let us assume the existence of $m$ Killing vectors $\partial_{\phi_i}$  of the metric $\gamma_{AB}$, where each $\phi_i$ is a $2\pi$ periodic coordinate on an axis of rotation of $\gamma_{AB}$.
Similar to before, we further assume that the vector $\nu^A$ can be written
\be\label{eq: nu vec}
\nu^A=\sum_{i=1}^{m}p_i(t,r)(\partial_{\phi_i})^A\,,
\ee
where $p_i(t,r)$ are arbitrary functions of $t$ and $r$. 
Since all the $\partial_{\phi_i}$'s mutually commute on $T{\cal S}$ their lifts are also Killing vectors of the metric $g_{\mu\nu}$.

\subsubsection*{Construction of Killing tensors}
We can also now assume that $\gamma_{AB}$ has  some particular  $n$-dimensional symmetry group $G(n)$ which is generated by some set of Killing vectors $(\xi_p)^A\in\mathfrak{g}(n)$ where $p\in S=\{1,\dots,n\}$ (not to be confused with the base space ${\cal S}$), of which the $\partial_{\phi_i}$'s generate an Abelian subgroup.
Let us write again the Lie algebra $\mathfrak{g}(n)$ of the symmetries as
\be
\llb \xi_p,\xi_q\rrb =f_{p q r}\xi_r\,.
\ee
Again, if $\mathfrak{g}$ is semi-simple and the $\{\xi_p\}$ form a basis then the structure constants are totally antisymmetric
\be 
f_{pqr}=f_{[pqr]}\,.
\ee

For example, $\gamma_{AB}$ could represent a metric of constant curvature, such as the $(d-2)$-sphere $S^{d-2}$, in which case the symmetry group would be $SO(d-2)$. 

Suppose now, that there exists a subset $I\subseteq S$ such that 
\be\label{eq: C s}
C_I=\sum_{p\in I\subseteq S} \xi_{p}\otimes\xi_{p}\,,
\ee
commutes with $\nu$. That is,
\be\label{eq: Cas phi com}
\llb C_I,\nu\rrb_{\SN}=\sum_i p_i(t,r)\llb C_I,\partial_{\phi_i} \rrb_{\SN}=2\sum_ip_i(t,r)\sum_{q\in I}  \xi_q\odot\llb\xi_q,\partial_{\phi_i}\rrb=0\,.
\ee
In the case that each of the functions $p_i$ are independent%
\footnote{
In the case that the metric is asymptotically flat or (A)dS this is equivalent to independent asymptotic angular momenta corresponding to the Killing vectors $\partial_{\phi_i}$ introduced Section \ref{sec: LT}.
}
this is equivalent to
\be
\llb C_I,\partial_{\phi_i} \rrb_{\SN}=0\,\quad \forall i\in\{1,\dots,m\}\,.
\ee
Now, the results of the previous section apply and we have immediately that $\hat{C}_I$ is a Killing tensor of $({\cal M},g)$ (here we again silently assume that all $\xi_p$ are `innate to' ${\cal S}$, that is $t$ and $r$ independent).

Moreover, if we have two such subsets $I,I'\subseteq S$, with possible overlap, i.e. $I\cap I'$ need not be empty, such that
\be
\llb C_I,\partial_{\phi_i} \rrb_{\SN}=0=\llb C_{I'},\partial_{\phi_i} \rrb_{\SN}\,\quad \forall i\in\{1,\dots,m\}\,,
\ee
Then if their SN bracket is non zero, by the Jacobi identity \eqref{eq: SN Jacobi}, we have a new non-trivial symmetry given by the following  rank 3 Killing tensor:
\be\label{rank3}
C_{II'}\equiv \llb C_{I}, C_{I'}\rrb_\SN\,, 
\ee
{which lifts to the full spacetime symmetry $\hat{C}_{II'}$.}
On the other hand, if this bracket vanishes, $C_I$ and $C_{I'}$ yield constants of motion on $\cal S$ and $\cal M$ which are in involution.
More explicitly, 
using  \eqref{eq: SN sym vecs} repeatedly, the SN bracket \eqref{rank3} can be written as
\begin{align}\label{KTsred}
    \llb C_I,C_{I'} \rrb_\SN^{ABC}
    &=2\sum_{q\in I'}\llb C_I,\xi_q\rrb_\SN^{(AB}\xi_q^{C)}
    =4\!\!\! \sum_{ p\in I\,,q\in I'}\xi_p^{(A}\xi_q^{B}\llb\xi_p,\xi_q \rrb^{C)}
    \nonumber\\
    &=4\!\!\! \sum\limits_{\substack{ p\in I,q\in I',
    \\r\in S}}
    f_{p q r} \,\xi_p^{(A}\xi_q^{B}\xi_r^{C)}
    \nonumber\\
    &=4 \sum_{r\in S}\Big(\sum\limits_{\substack{ p\in I\backslash I'\\
    q\in I'\backslash I}} +\sum\limits_{\substack{ p\in I\cap I'\\
    q\in I'\backslash I}}+\sum\limits_{\substack{ p\in I\backslash I'\\
    q\in I'\cap I}}\Big)
    f_{p q r} \,\xi_p^{(A}\xi_q^{B}\xi_r^{C)}
    \,.
\end{align}
Notice, that due to the automatic antisymmetry of the structure constants on the first two indices, the sum on $p$ and $q$ must be over distinct ranges. 
Moreover, if we have a semi-simple Lie algebra for which $\xi_p$ form a basis then 
the ranges of $p$, $q$, and $r$ all have to  be all distinct.
Of course, this relation continues to hold at the level of the spacetime, i.e.
\begin{align}\label{KTfull}
[{\hat C}_{I},{\hat C}_{I'}]_{\SN}^{\mu\nu\rho}\partial_\mu\partial_\nu\partial_\rho
&=4 \sum_{r\in S}\Big(\sum\limits_{\substack{ p\in I\backslash I'\\
    q\in I'\backslash I}} +\sum\limits_{\substack{ p\in I\cap I'\\
    q\in I'\backslash I}}+\sum\limits_{\substack{ p\in I\backslash I'\\
    q\in I'\cap I}}\Big)f_{p q r} \,\hat{\xi}_p^{(\mu}\hat{\xi}_q^{\nu}\hat{\xi}_r^{\rho)}\partial_\mu\partial_\nu\partial_\rho
    \nonumber\\
    &=4 \sum_{r\in S}\Big(\sum\limits_{\substack{ p\in I\backslash I'\\
    q\in I'\backslash I}} +\sum\limits_{\substack{ p\in I\cap I'\\
    q\in I'\backslash I}}+\sum\limits_{\substack{ p\in I\backslash I'\\
    q\in I'\cap I}}\Big)f_{p q r} \,\xi_p^{(A}\xi_q^{B}\xi_r^{C)}\partial_A\partial_B\partial_C\,.
\end{align}
However, whereas \eqref{KTsred} is a reducible Killing tensor on the base space (simply written as a linear combination of symmetrized products of various 3 Killing vectors), \eqref{KTfull} is an irreducible Killing tensor of the full spacetime, provided not all Killing vectors $\xi$ lift to Killing vectors on the spacetime.

\subsubsection*{Mutual commutativity and geodesic integrability}

If we want a {\em completely integrable} geodesic system, then we need a set of $(d-1-m)$ mutually commuting Killing tensors to complement $m$ axial Killing vectors, $\partial_{\phi_i}$ plus the inverse metric $g^{-1}$. 
{If, in addition, the spacetime is also stationary, an extra timelike Killing vector $\partial_t$ is present, and we need} one less Killing tensor lifted from $\cal S$.

One way that a set of {\em mutually commuting} Killing tensors of $\cal M$ could arise is if we have a collection of nested subsets $I_1\subset I_2\subset...\subset I_N$, each of which generate a Lie subalgebra, such that each
\be\label{eq: Cas}
C_{I_i}=\sum_{p\in I_i\subseteq S} \xi_{p}\otimes\xi_{p}=\sum_{p\in I_i} \xi_{p}\otimes\xi_{p}\,,
\ee
commute with all of the rotational Killing vectors,
\be
\llb C_{I_i},\partial_{\phi_k} \rrb_{\SN}=0\,\quad \forall k\,.
\ee
From \eqref{KTsred} it is clear that any two such Killing tensors $C_{I_i}$ and $C_{I_j}$  will not necessarily commute with each other. Now, since either $I_i\subset I_j$, or  $I_j\subset I_i$, if $i<j$, or $j<i$ respectively, the first term in the sum in \eqref{KTsred} will be empty. However, one of the last two terms will survive: e.g. if $I_i\subset I_j$,
\be\label{eq: Com obstruction}
\llb C_{I_i}, C_{I_j}\rrb=4\sum\limits_{\substack{ p,r\in I_i\\
    q\in I_j\backslash I_i}}f_{p q r} \,\xi_p\odot\xi_q\odot\xi_r\,,
\ee
where the summation range of $r$ is restricted to $I_i$ because the set generates a Lie subalgebra and so is closed.

If, in addition, the Killing vectors that these subsets label form \emph{semi-simple} Lie subalgebras, i.e. $\xi_p,\xi_q, \xi_r$ for $p,q,r\in I_i$
\be
\llb \xi_p,\xi_q\rrb =\tilde{f}_{pqr}\xi_r\,,\quad \tilde{f}_{[pqr]}=\tilde{f}_{pqr}\,,
\ee
then \eqref{eq: Com obstruction} vanishes (similarly to \eqref{eq: SN brack Cas}) and the quadratic Killing tensor $C_{I_i}$ commutes with each $\xi_p$
\be
\llb C_{I_i}, \xi_p\rrb_\SN=0\,,
\ee
hence they are Casimirs of the subalgebra and we have the desired commutativity.%
\footnote{Essentially this construction is related to the Noetherian property of the universal enveloping algebra (see section 7 of ~\cite{robson1988noncommutative}). For example, in $SO(d-2)$ we have the canonical chain~\cite{iachello2006lie} 
$SO(d-2)\supset SO(d-3)\dots \supset SO(2)$ and so the nested subsets correspond to labelling rotation subgroups.
It may be of interest to make such statements precise, at least for the case where $\gamma$ is maximally symmetric.
}

Finally, the results of the preceding calculations (in particular \eqref{eq: Lifted bracket}) show that the lift, $\hat{C}_{I_i}$, of the Casimirs to the full spacetime ${\cal M}$ form a set of mutually commuting Killing tensors of $g_{\mu\nu}$. 
That is,
\be
[{\hat C}_{I_i},g^{-1}]^{\mu\nu\rho}_{\SN}=0\,,\quad [{\hat C}_{I_i},{\hat C}_{I_j}]^{\mu\nu\rho}_{\SN}=0\,.
\ee
Furthermore, each ${\hat C}_S$ is irreducible to Killing vectors provided one removes any set of the axial Killing vectors, $\partial_{\phi_i}$, which may have appeared in the sum \eqref{eq: Cas}. 
In fact one could already quotient out this subgroup in \eqref{eq: Cas} because all the $\partial_{\phi_i}$'s mutually commute and we are only interested in additional symmetries.

\subsection{Growing tower of higher rank Killing tensors}

In the previous we have shown as to how the reducible `Casimirs' on ${\cal S}$ can give rise to irreducible rank 2 Killing tensors on ${\cal M}$, and that, in their turn, these may generate new (possibly irreducible) rank 3 Killing tensors on ${\cal M}$ via the SN bracket. Starting from these objects and applying the SN bracket iteratively, one can, at least in principle, generate a tower of (possibly irreducible) growing rank Killing tensors.

{For example, we can generate the following rank 4 Killing tensor (applying the Jacobi identity \eqref{eq: SN Jacobi}):}
\ba
 \Big\llb\llb C_I,C_{I'} \rrb,C_{I''}\Big\rrb^{ABCD}_{\SN}
 &=&\qquad\sum\limits_{\substack{p\in I, q\in I'\\
                                 s\in I'', r\in S}}
                                 \bigg[ 12f_{pqr}\xi^{(M}\xi^A_q\xi_r^B\partial_M(\xi_s^C\xi_s^{D)}) \qquad\nn\\
 &&\qquad \qquad-8\xi_s^M\xi^{(A}_sf_{pqr}\partial_M(\xi^B_p\xi^C_q\xi^{D)}_r)\bigg]\\
  &=&\, -8\sum\limits_{\substack{p\in I, q\in I',s\in I''\\
                r,t\in S}}
                \bigg[f_{pqr}f_{spt}\xi^{(A}_s\xi^B_t\xi^C_q\xi^{D)}_r+\nn\\
&&\qquad \quad+f_{pqr}f_{sqt}\xi^{(A}_s\xi^B_p\xi^C_t\xi^{D)}_r
+f_{pqr}f_{srt}\xi^{(A}_s\xi^B_p\xi^C_q\xi^{D)}_t\bigg]\,.\quad
\ea
As before, {the summation ranges will reduce when $\mathfrak{g}$ is a semi-simple Lie algebra (totally antisymmetric structure constants); however, the resulting expression is not particularly enlightening and so we do not reproduce it here.}

More generally, taking repeated SN brackets we can go to arbitrary rank $k+1$ as follows:
\begin{align}\label{eq: higher KTs}
    \Big\llb\big\llb\cdots\llb C_{I_1},C_{I_2} \rrb,\cdots\big\rrb,C_{I_k}\Big\rrb&=-(-2)^k\prod_{i=0}^{k-2}\sum_{\substack{s_j\in I_j\\ r_j\in S\\ t_i\in\{s_1,\dots,s_{i+1},r_1,\dots,r_i\}\backslash\\\{t_1,\dots,t_{i-1}\}}}f_{s_{i+2}t_ir_{i+1}}\bigodot_{\substack{u\in \{s_1,\dots,s_k,\\r_1,\dots,r_{k-1}\}\backslash\\\{t_1,\dots,t_{k-2}\}}}\xi_u\,,
\end{align}
where again the summation range of the first two indices $s_{i+2}$, $t_i$ in each structure $f_{s_{i+2}t_i r_i}$ constant must be disjoint and, in particular, when the algebra is semi-simple all three summation ranges must be disjoint (when appearing in combination with the symmetrized vector product $\xi_{s_{i+2}}\odot\xi_{t_i}$ or $\xi_{s_{i+2}}\odot\xi_{t_i}\odot\xi_{r_{i+1}}$ respectively). 
However, similar to {the last line of \eqref{KTsred}} each of the terms in the product will have a different collection of (disjoint) summation ranges. 
Hence, it is not {\em obvious} that {any cancellations cause this nested commutator to terminate.} 
Furthermore, as the summation terms are distinct  {\eqref{eq: higher KTs}} is not reducible to the Killing tensors in \eqref{eq: Cas} which involve a summation over products of two identical vectors. 
Together, this hints that these higher rank Killing tensors are irreducible.

On the other hand, the process in \eqref{eq: higher KTs} will not generate an infinite set of functionally independent constants of geodesic motion since the commutators are determined from the structure constants of the group.
In particular, for geodesic motion confined to $\cal S$ with dimension $d-2$ there can only be $2d-5$ functionally independent constants of motion (corresponding to a maximally superintegrable system) \cite{Miller_2013}.
Furthermore, the maximum number of independent rank-$p$ Killing tensors (including Killing vectors), achieved when ${\cal S}$ is maximally symmetric, is (see e.g.~\cite{Houri:2017tlk} for an extended discussion and the original references~\cite{barbance1973tenseurs,DeLong1982,Takeuchi1983,Thompson1986}): 
\be \label{eq: Kmax}
k_{\max}=\frac{1}{d-2}
\begin{pmatrix}
(d-2)+p\\
p+1
\end{pmatrix}
\begin{pmatrix}
(d-2)+p-1\\
p
\end{pmatrix} \,.
\ee 
This gives a bound on the number of Killing tensors of rank $p=k+1$ generated by $k$ subsets $I_1, \dots ,I_k$.

\section{Generalized Lense--Thirring spacetimes}\label{sec: LT}
We will now apply our formalism to a physically motivated example, namely the class of generalized Lense--Thirring spacetimes~\cite{Gray:2021toe,Gray:2021roq,Gray:2024qys}, which are an off-shell (i.e. no field equations imposed) class of spacetimes well suited to describing slowly rotating black holes in many different theories; as discussed at the end of this section, they  also represent an exact rotating black hole solution in the EMDA theory.

\subsection{ Introducing the metric }

The generalized Lense--Thirring metric is stationary and axially symmetric and reads%
\footnote{
As discussed in the previous section, the assumption of stationarity is not necessary, and one could consider time dependent metric functions $N, f$ and $p_i$. In such a case, the same symmetries except for $\partial_t$ would arise.
}
\be\label{LTHDimproved}
ds^2=-Nfdt^2+\frac{dr^2}{f}+r^2\sum_{i=1}^m \mu_i^2\Bigl(d\phi_i+p_i dt\Bigr)^2+r^2\sum_{i=1}^{m+\epsilon}d\mu_i^2\;.
\ee
Here, as before, $m=[\frac{d-1}{2}]$ and now $\epsilon=1, 0$ for even, odd dimensions $d=2m+\epsilon+1$, respectively. 
The metric functions $f, N, p_i$ are functions of the radial coordinate $r$, and the `angular' coordinates $\mu_i$ are constrained as follows: 
\be\label{constraint}
\sum_{i=1}^{m+\epsilon} \mu_i^2=1\,.
\ee
Moreover, in \cite{Gray:2021roq}, it was useful to define 
\be\label{gen pi}
p_i(r)=\frac{{\sum_{j=1}^mp_{ij}(r)a_j}}{r^2}\,, 
\ee
to explicitly introduce the corresponding rotation parameters $a_i$. 

The above introduced spacetime is off-shell, meaning that no field equations have been imposed. However, by specifying the metric functions $f, N, p_i$ one can obtain approximate (slowly rotating) black hole solutions in various theories.
For example, the vacuum Einstein gravity slowly rotating black solution in all dimensions is recovered upon setting 
\be \label{Einstein}
N=1\,,\quad f=1-\Bigl(\frac{r_+}{r}\Bigr)^{d-3}\,,\quad p_i=\frac{a_i}{r^2}(f-1)\,, 
\ee 
where the parameter $r_+$ is related to the black hole mass and parameters $a_i$ are its rotation parameters (see \cite{Gray:2021roq} for more such examples). 

{More generally, let us assume that the above spacetime describes a black hole solution. In that case the black hole horizon is located at $r_+$, given by the due root of $f(r_+)=0$. Such a horizon is then a Killing horizon} generated by the following Killing vector:
\be\label{killinggen}
\xi=\partial_t+\sum_{i=1}^m \Omega^i_+ \partial_{\phi_i}\,,
\quad 
\Omega^i_+\equiv -p_i(r_+)\,.
\ee
The spacetime then features an ergoregion, a region outside of the horizon where the Killing vector $\partial_t$ has negative norm; $\Omega_i$ here play the role of angular velocities of the horizon. 
Moreover, one can impose leading order asymptotic flat or (A)dS fall-off conditions for large $r$:
\be
N=1+O(r^{-k_N})\,,\: f=-\frac{2\Lambda r^2}{(d-1)(d-2)}  +1-\frac{2 m}{r^{d-3}}+O(r^{-k_f})\,,\: p_i=\frac{\sum_{j=1}^m p^{0}_{ij}a_j}{r^{d-5}}+O(r^{-k_{p_i}})\,,
\ee
where $\Lambda$ is the cosmological constant,  $p_{ij}^0$ is the leading order term of $p_{ij}(r)$ in \eqref{gen pi}, and $\{k_N,k_f,k_{p_i}\}$ are positive integers.%
\footnote{In principle, one could choose a more general, e.g. polyhomogeneous~\cite{Chrusciel:1993hx}, asymptotic expansion but that is not required here.
}
Then, to each of the Killing vectors one can associate the generalized Komar charge~\cite{Bazanski:1990qd,Kastor:2008xb}:
\be
{\cal Q}[\zeta]\propto\int_{S^{d-2}_\infty}\star\left( d\zeta-\frac{4\Lambda}{d-2}\omega^\zeta\right)\,,
\ee
where $\omega^\zeta$ is a Killing co-potential $\nabla^a\omega^\zeta_{ab}=\zeta_b$. In this case the mass and angular momenta of the spacetime are
\be
M\equiv {\cal Q}[\partial_t]\propto m\,,\quad  J_i\equiv {\cal Q}[\partial_{\phi_i}]\propto \sum_{j=1}^m p^{0}_{ij}a_j\,.
\ee
Thus, the requirement for independent asymptotic angular momenta is equivalent to the requirement for independent $p_i$ functions.

Irrespective of its physical interpretation, it was shown in \cite{Baines:2020unr, Baines:2021qaw, Gray:2021toe, Gray:2021roq, Sadeghian:2022ihp, Gray:2024qys} that the above generalized Lense--Thirring spacetime \eqref{LTHDimproved} possesses many remarkable properties. 
Namely, it can be cast in the Painlev{\'e}--Gullstrand form and admits  arapidly growing tower of (increasing rank) Killing tensors which guarantee separability of the Hamilton--Jacobi equation for geodesics and the Klein--Gordon equation for scalar fields (see~\cite{Sadeghian:2022ihp} and \cite{Gray:2024qys} for their respective explicit separation). 
Our goal here is to demystify the appearance of such hidden symmetries.

\subsection{Rank-2 Killing tensors and their brackets}

To start with, observe that the metric \eqref{LTHDimproved} is of the form \eqref{Generic} with the vector $\nu^A$ given by \eqref{eq: nu vec}. 
Clearly, we have the Killing vectors of the full metric $\partial_t$ and $\partial_{\phi_i}$, while the metric on $\cal S$ is 
\begin{equation}
    \gamma_{AB} dx^A dx^B= r^2\bigg(\sum_{i=1}^{m+\epsilon}d\mu_i^2+ \sum_{i=1}^m \mu_i^2 d\phi_i^2\bigg)\,,
\end{equation}
which is just the metric on the $(d-2)$-sphere. 
Naturally, the symmetry group is simply $SO(d-1)=SO(2m+\epsilon)$ which becomes manifest in the coordinates on the $i$-th two-plane, namely~\cite{Myers:1986un,Myers:2011yc}:
\begin{align}
x_i&=\mu_i\cos \phi_i\,,\quad y_i=\mu_i \sin \phi_i\,,\quad 1=\sum_{i=1}^{m+\epsilon}x_i^2+y_i^2\,, \quad y_{m+1}=0\,,
\end{align}
upon which the metric on $\cal S$ becomes
\be
\gamma_{AB} dx^A dx^B= r^2\bigg(\epsilon dx_{m+\epsilon}^2+\sum_{i=1}^{m}\left[dx_i^2+dy_i^2\right]\bigg)\,.
\ee

In these coordinate the symmetry algebra is generated by the following Killing vectors for $i,j\leq m+\epsilon$:%
\footnote{In these coordinates it is clear that if $x_{m+\epsilon}, y_{m+\epsilon}$ were not constrained, we would have just a flat metric, with the usual Killing vectors of $(2m+\epsilon)$-dimensional Euclidean space, i.e. rotations and translations. 
However, due to the constraint we are left with only the rotations in \eqref{eq: L KVs}.
}
\begin{align}\label{eq: L KVs}
L_{x_i x_j}&=x_i\partial_{x_j}-x_j\partial_{x_i}
\nonumber\\
&=\cos\phi_i\cos\phi_j(\mu_i\partial_{\mu_j}-\mu_j\partial_{\mu_i})-\cos\phi_i\sin\phi_j\frac{\mu_j}{\mu_i}\partial_{\phi_i}+\cos\phi_j\sin\phi_i\frac{\mu_i}{\mu_j}\partial_{\phi_j}
\,,
\nonumber\\
L_{x_i y_j}&=x_i\partial_{y_j}-y_j\partial_{x_i}
\nonumber\\
&=\cos\phi_i\sin\phi_j(\mu_i\partial_{\mu_j}-\mu_j\partial_{\mu_i})+\cos\phi_i\cos\phi_j\frac{\mu_i}{\mu_j}\partial_{\phi_j}+\sin\phi_i\sin\phi_j\frac{\mu_j}{\mu_i}\partial_{\phi_i}
\,,
\nonumber\\
L_{y_i y_j}&=y_i\partial_{y_j}-y_j\partial_{y_i}
\nonumber\\
&=\sin\phi_i\sin\phi_j(\mu_i\partial_{\mu_j}-\mu_j\partial_{\mu_i})
+\sin\phi_i\cos\phi_j\frac{\mu_j}{\mu_i}\partial_{\phi_i}-\sin\phi_j\cos\phi_i\frac{\mu_i}{\mu_j}\partial_{\phi_j}
\,,
\end{align}
where it should be understood that  $\partial_{\mu_{m+\epsilon}}=0=\partial_{\phi_{m+1}}$.
Notice $L_{x_iy_i}=\partial_{\phi_i}$ and $L_{y_iy_{m+1}}=0=L_{x_i y_{m+1}}$.%
\footnote{Thus, in all of the following expressions with sums it should be understood that $y_{m+1}$ does not appear.}
In total this gives the required number, $\frac{1}{2}(2m+\epsilon)(2m+\epsilon-1)$, of  generators.

If $I,J,K,L$ stand for any of $x_i,y_j$ (except $y_{m+1}=0$) then we can express the algebra as simply
\be
\llb L_{IJ}, L_{KL}\rrb =\delta_{I L} L_{JK}+\delta_{JK} L_{IL} -\delta_{IK} L_{JL}-\delta_{JL}L_{IK}\equiv f_{IJKLMN}L_{MN}\,.
\label{eq: LT struct} 
\ee
This is exactly the usual algebra of $\mathfrak{so}(2m+\epsilon)$ with structure constants in this basis
\begin{align}
    f_{IJKLMN}=\frac{1}{2}(
    &\delta_{IL}\delta_{JM}\delta_{KN}+\delta_{J K}\delta_{I M}\delta_{LN}-\delta_{I K}\delta_{J M}\delta_{KN}-\delta_{JL}\delta_{IM}\delta_{KN}
    \nonumber\\
    &-\delta_{IL}\delta_{JN}\delta_{KM}-\delta_{J K}\delta_{I N}\delta_{LM}+\delta_{I K}\delta_{J N}\delta_{KM}+\delta_{JL}\delta_{IN}\delta_{KM})
\,.\label{eq: LT struct explicit} 
\end{align}

We can then introduce the Casimirs for each plane 
\begin{align}\label{eq: LT Cass}
C_{ij}&=\frac{1}{2}\sum\limits_{\substack{ I\in\{x_i,y_i\}\\
J\in\{x_j,y_j\}}}L_{IJ}\odot L_{IJ}
\nonumber\\
&=\frac{1}{2}\left( L_{x_i x_j}\odot L_{x_i x_j}+L_{x_i y_j}\odot L_{x_i y_j}+L_{y_i x_j}\odot L_{y_i x_j}+L_{y_i y_j}\odot L_{y_i y_j}\right)
\nonumber\\
    &=\frac{1}{2}\bigg((\mu_i\partial_{\mu_j}-\mu_j\partial_{\mu_i})^2 +\Big(\frac{\mu_i}{\mu_j}\partial_{\phi_j}+\frac{\mu_j}{\mu_i}\partial_{\phi_i}\Big)^2\bigg)\,.
\end{align}
We call them Casimirs here because they {SN commute} with any Killing vector in a different two-plane to $i$ and $j$, i.e. $\llb C_{ij}, L_{kl}\rrb_{\SN}=0$ for $\{i,j\}\neq \{k,l\}$. 
This follows immediately from the Kronecker deltas in \eqref{eq: LT struct}.
Moreover the Casimirs \eqref{eq: LT Cass} also commute with all $\partial_{\phi_k}=L_{x_k,y_k}$.
Namely, for all $i,j,k$ we have (no sum)
\begin{align}\label{eq: Cass phi comm LT}
   \llb C_{ij},\partial_{\phi_k}\rrb_\SN&=\frac{1}{2}
   \sum\limits_{\substack{ I\in\{x_i,y_i\}\\ J\in\{x_j,y_j\} }}
   \llb L_{IJ}\odot L_{IJ}, L_{x_k,y_k}\rrb_{\SN}
   =\sum\limits_{\substack{ I\in\{x_i,y_i\}\\ J\in\{x_j,y_j\} }}\llb L_{IJ}, L_{x_k,y_k}\rrb\odot L_{IJ}
   \nonumber\\
   &=\sum\limits_{\substack{ I\in\{x_i,y_i\}\\ J\in\{x_j,y_j\} }}\left( \delta_{I y_k} L_{J x_k} 
   +\delta_{j x_k} L_{I y_k}- \delta_{I x_k} L_{J y_k}- \delta_{J y_k} L_{I x_k}
   \right)\odot L_{IJ}
   \nonumber\\
   &=\delta_{ik} \sum_{J\in\{x_j,y_j\}}   
   ( L_{J x_k}\odot L_{y_k J}- L_{J y_k}\odot L_{x_k J})
   \nonumber\\
   &+ 
   \delta_{jk} 
   \sum_{I\in\{x_i,y_i\}} (L_{I y_k}\odot L_{I x_k}
   - L_{I y_k}\odot L_{x_k I})
    =0
   \,.
\end{align}

Thus, from our previous discussion and in particular \eqref{eq: Lifted bracket}, we see that each $C_{ij}$ lifts to a Killing tensor of the full spacetime 
\be \label{eq: LT KTs} 
{\hat C}_{ij}=\frac{1}{2}\bigg((\mu_i\partial_{\mu_j}-\mu_j\partial_{\mu_i})^2 +\Big(\frac{\mu_i}{\mu_j}\partial_{\phi_j}+\frac{\mu_j}{\mu_i}\partial_{\phi_i}\Big)^2\bigg)\,.
\ee
It should be observed that not all of these are independent since $C_{ij}=C_{ji}$. 
Moreover, $C_{ii}=2\partial_{\phi_i}\otimes\partial_{\phi_i}$ so these are only irreducible Killing tensors on $\cal M$; for $m+\epsilon\geq i>j>1$. 
Therefore, there are $\frac{1}{2}(m+\epsilon)(m+\epsilon-1)$ irreducible Killing tensors. 
We note that these Killing tensors do not all commute with each other. 
Using \eqref{KTsred} and \eqref{eq: LT struct} we have
\begin{align}\label{eq: Cass comm LT}
    \llb C_{ij},C_{kl}\rrb_\SN
    &=\sum\limits_{\scriptstyle \substack{ I\in\{x_i,y_i\}\\ J\in\{x_j,y_j\} 
    }}
    \sum\limits_{\scriptstyle \substack{K\in\{x_k,y_k\} \\ L\in\{x_l,y_l\}
     }}
    \left( \delta_{I L} L_{JK}+ \delta_{J K} L_{IL} - \delta_{I K} L_{JL}
     -\delta_{J L} L_{IK}\right) \odot L_{IJ}\odot L_{KL}
    \nonumber\\
    &=\delta_{i l}\sum\limits_{\scriptstyle \substack{ I\in\{x_i,y_i\}\\ J\in\{x_j,y_j\} 
    \\
     K\in\{x_k,y_k\}}}
    L_{JK} \odot  L_{IJ}\odot  L_{KI}
    +\delta_{j k}\sum\limits_{\scriptstyle \substack{ I\in\{x_i,y_i\}\\ J\in\{x_j,y_j\} 
    \\
     L\in\{x_l,y_l\}}}
    L_{IL}\odot L_{IJ} \odot L_{JL}
    \nonumber\\
    &-\delta_{i k}\sum\limits_{\scriptstyle \substack{ I\in\{x_i,y_i\}\\ J\in\{x_j,y_j\} 
    \\
     L\in\{x_l,y_l\}}}
    L_{JL} \odot  L_{IJ}\odot  L_{IL}
    -\delta_{j l}\sum\limits_{\scriptstyle \substack{ I\in\{x_i,y_i\}\\ J\in\{x_j,y_j\} 
    \\
     K\in\{x_k,y_k\}}}
    L_{IK}\odot L_{IJ} \odot L_{KJ}
    \nonumber\\
    &=
    (\delta_{j l}-\delta_{il})\sum\limits_{\scriptstyle \substack{ I\in\{x_i,y_i\}\\ J\in\{x_j,y_j\} 
    \\
     K\in\{x_k,y_k\}}}
     \!\!\!
     L_{IJ}\odot  L_{IK} \odot L_{JK} 
   +(\delta_{j k}-\delta_{i k})
    \sum\limits_{\scriptstyle \substack{ I\in\{x_i,y_i\}\\ J\in\{x_j,y_j\} 
    \\
     L\in\{x_l,y_l\}}}
     \!\!\!
     L_{IJ} \odot L_{IL}\odot L_{JL}
    \,.
\end{align}
We can observe the following consequences of \eqref{eq: Cass comm LT}: 
\begin{enumerate}
    \item As expected, when $l=k$ or when $i=j$ the commutator vanishes corresponding to the situation that either $C_{ii}=2\partial_{\phi_i}^2 $ or $C_{kk}=2\partial_{\phi_k}^2$ and hence the bracket must vanish by \eqref{eq: Cass phi comm LT}. 

    \item The expression is symmetric on the swap of indices $i\leftrightarrow j$ or $k\leftrightarrow l$ as required by \eqref{eq: LT Cass}.
    
    \item Since the SN bracket is antisymmetric, the right hand side of \eqref{eq: Cass comm LT} is antisymmetric under the swap of pairs of indices $\{ij\}\leftrightarrow\{kl\}$. This is not immediately obvious but requires relabelling using the Kronecker deltas.
    
    \item For $i\neq j$ and $k\neq l$ this bracket does not vanish if one of the indices equals each other, as then one of the Kronecker deltas will be nonzero. 
    Thus, when $\{i,j,k,l\}$ are all distinct \eqref{eq: LT Cass} gives a set of new non-trivial (i.e. irreducible) constants of motion which are mutually commuting, i.e. in involution.
\end{enumerate}

Point 4  gives a total of $m+\epsilon-2$ Casimirs, corresponding to the distinct pairs, which are mutually commuting. We can add to this
\be
C=\sum_{i,j=1}^{m+\epsilon}C_{ij}=r^2\gamma^{-1}\,,
\ee
which commutes with all other $C_{kl}$'s. Thus giving $m+\epsilon-1$ mutually commuting Killing tensors of $\cal S$. 
In addition we have the $m$ Killing vectors $\partial_{\phi_i}$ and thus we have explicitly obtained the $d-3=2m+\epsilon-1$ symmetries necessary for the complete integrability of motion in the $(d-2)$-sphere.
Moreover the motion in the full spacetime $\cal M$ is also completely integrable as we also have the full metric $g$ and timelike Killing vector $\partial_t$.

 We now make contact with the previous work on the generalized LT spacetimes~\cite{Gray:2021roq,Gray:2021toe,Sadeghian:2022ihp}.

\subsection{Comparison with previous results}

\subsubsection*{Alternative expressions for Killing tensors}

To begin, the Killing tensors ${\hat C}_{ij}$, \eqref{eq: LT KTs}, were first 
written down in~\cite{Sadeghian:2022ihp} in response to the works~\cite{Gray:2021roq,Gray:2021toe}.
However, they were described in terms of the $\mathfrak{u}(m)$ subalgebra with generators (see ~\cite{Galajinsky:2013mla,Vasudevan:2004ca})
\begin{align}
    \xi_{ij}&=L_{x_ix_j}+L_{y_iy_j}=x_i\partial_{x_j}-x_j\partial_{ x_i}+y_i\partial_{y_j}-y_j\partial_{y_i}
      \nonumber\\
        &=\cos(\phi_i-\phi_j)(\mu_i\partial_{\mu_j}-\mu_j\partial_{\mu_i})+\sin(\phi_i-\phi_j)\Big(\frac{\mu_i}{\mu_j}\partial_{\phi_j}+\frac{\mu_j}{\mu_i}\partial_{\phi_i}\Big)
    \,,\label{eq: xi KV}
    \\
    \rho_{ij}&=L_{x_i y_j}+L_{x_j y_i}=x_i\partial_{y_j}-y_j\partial_{x_i}
    +x_j\partial_{y_i}-y_i\partial_{x_j}
    \nonumber\\
    &=-\sin(\phi_i-\phi_j)(\mu_i\partial_{\mu_j}-\mu_j\partial_{\mu_i})+\cos(\phi_i-\phi_j)\Big(\frac{\mu_i}{\mu_j}\partial_{\phi_j}+\frac{\mu_j}{\mu_i}\partial_{\phi_i}\Big)
    \,,\label{eq: rho KV}
\end{align}
for $i,j\in\{1,\dots m\}$. 
Explicitly these generators satisfy the following $\mathfrak{u}(m)$ sub-algebra:
\begin{align}\label{eq: u(m) algebra}
  [\xi_{ij},\xi_{kl}]&=  (\delta_{jk}\xi_{il}+\delta_{il}\xi_{jk}-\delta_{ik}\xi_{jl}-\delta_{jl}\xi_{ik})\,,
  \nonumber\\
  [\rho_{ij},\rho_{kl}]&=  ( -\delta_{jk}\xi_{il}-\delta_{ik}\xi_{jl}-\delta_{il}\xi_{jk}-\delta_{jl}\xi_{ik})\,,\quad
  \nonumber\\
   [\xi_{ij},\rho_{kl}]&=  (\delta_{jk}\rho_{il}+\delta_{jl}\rho_{ik}-\delta_{ik}\rho_{jl}-\delta_{il}\rho_{jk})\,.
\end{align}
Moreover, the square of these generators gives the same Casimir in \eqref{eq: LT Cass}, i.e.
\be
C_{ij}=\frac{1}{2}(\xi_{ij}\otimes\xi_{ij} +\rho_{ij}\otimes \rho_{ij})\,,
\ee
but they do not cover the $C_{i\, m+\epsilon}$ cases. 
Furthermore, the relationship between the symmetries of the transverse space $\cal S$ and the full spacetime, as in \eqref{eq: Lifted bracket}, was not observed. 

Next, as previously shown in~\cite{Gray:2021roq,Gray:2021toe} up to $d=13$, the Lense--Thirring spacetimes \eqref{LTHDimproved} admit the following Killing tensors explicitly written as:
\begin{align}\label{KTs}
K^{(I)}=&\sum\limits_{i\not\in I}^{m-1+\epsilon}\bigg[\bigr(1-\mu_i^2-\!\sum_{j\in I}\mu_j^2\bigr)(\partial_{\mu_i})^2 -2\!\!\sum_{j\not\in I\cup\{i\}}\!\! \mu_i\mu_j\,\partial_{\mu_i}\partial_{\mu_j} \bigg] \nonumber\\
&\qquad +\sum\limits_{i\not\in I}^{m}\bigg[\frac{1-\sum_{j\in I}\mu^2_j}{\mu_i^2}(\partial_{\phi_i})^2\bigg] \,.
\end{align}
where we define the set $S=\{1,..,m\}$ (not to be confused with the base space ${\cal S}$) and let $I\in P(S)$ where $P(S)$ is the power set of $S$, i.e. the set of all subsets $I\subset S$. 
Again not all of these are independent, in particular $\sum_{i=0}^{m-3}\binom{m}{i}$ of these are reducible, leaving
\be
k=\sum_{i=0}^{m-2+\epsilon}\binom{m}{i}-\sum_{i=0}^{m-3} \binom{m}{i} =\frac{1}{2} (m+\epsilon)(m+\epsilon-1)
\ee
irreducible rank-2 Killing tensors in $d$ dimensions.%
\footnote{Previously the following equivalent expression was stated (for $\epsilon=0,1$) in \cite{Gray:2021roq,Gray:2021toe}:
\be 
k=\frac{1}{2}m(m-1+2\epsilon)\,.
\ee
However, the current form makes the structure clearer.}
The question then remains: what is the relationship between these two expressions for the Killing tensors?

Since the number of independent Killing tensors, $\hat{C}_{ij}$ and $K^{(I)}$, is the same, it is perhaps not surprising that we can express them in terms of each other. 
Namely, using the constraint \eqref{constraint} we can write
\begin{align}\label{eq: KT equivalence}
 K^{(I)}=  &\sum\limits_{i\not\in I}^{m+\epsilon}\sum\limits_{j\not\in I\cup\{i\}}^{m+\epsilon}
 \bigg[ \mu_j^2(\partial_{\mu_i})^2 -2\mu_i\mu_j\,\partial_{\mu_i}\partial_{\mu_j} \bigg]
 +\sum\limits_{i\not\in I}^{m+\epsilon}\sum\limits_{j\not\in I\cup\{i\}}^{m+\epsilon}\frac{\mu^2_j}{\mu_i^2}(\partial_{\phi_i})^2 
 \nonumber\\
 &=\frac{1}{2}\left(\sum\limits_{i\not\in I}^{m+\epsilon}\sum\limits_{j\not\in I\cup\{i\}}^{m+\epsilon}
 \big( \mu_i\partial_{\mu_j}-\mu_j\partial_{\mu_i} \big)^2
 +
 \sum\limits_{i\not\in I}^{m+\epsilon}\sum\limits_{j\not\in I\cup\{i\}}^{m+\epsilon}
 \left[\left(\frac{\mu_j}{\mu_i}\partial_{\phi_i}+ \frac{\mu_j}{\mu_i}\partial_{\phi_i}\right)^2 -2\partial_{\phi_i}\partial_{\phi_j}\right]\right)
 \nonumber\\
 &=\sum\limits_{i\not\in I}^{m+\epsilon}\sum\limits_{j\not\in I\cup\{i\}}^{m+\epsilon}\left(\hat{C}_{ij}-2\partial_{\phi_i}\partial_{\phi_j}\right)\,,
\end{align}
where we have used the symmetry on the summed indices and it should be understood that, as above, $\partial_{\phi_{m+1}}=0=\partial_{\mu_{m+\epsilon}}$, and $\mu_{m+\epsilon}$ is given by the constraint \eqref{constraint}.

Therefore, up the reducible product of Killing vectors we see that $K^{(I)}$ is given by a sum over the Killing tensors lifted from the $SO(2m+\epsilon)$ symmetry of the transverse space $\cal S$. 
Moreover, the previously obtained expressions \eqref{KTs} are therefore valid in every dimension.

\subsubsection*{Commutation relations}

 Previously it was verified in ~\cite{Gray:2021roq,Gray:2021toe} up to $d=13$, that the SN bracket of any two Killing tensors \eqref{KTs} vanishes if the intersection of the two label sets equals the first. That is,
\be\label{eq: KT Comms LT}
[K^{(I_1)}, K^{(I_2)}]_\SN=0\,, \quad \text{ iff $I_1\subset I_2$ or $I_2\subset I_1$ .}
\ee
In the framework of this paper, such a commutation relation seems to follow from the general discussion in Section \ref{sec: Gen ax symm}. 
However, we now seek to prove this relation using the explicit algebra of Killing vectors of $\cal S$. From \eqref{eq: KT equivalence} and \eqref{eq: Cass comm LT} we have
\begin{align}\label{eq: KT Comm LT}
    [K^{(I_1)}, K^{(I_2)}]_\SN&=\sum\limits_{
    \substack{i\not\in I_1\\
    j\not\in I_1\cup\{i\}
    }}^{m+\epsilon}
    \sum\limits_{
    \substack{k\not\in I_2\\
    l\not\in I_2\cup\{k\}
    }}^{m+\epsilon}
    [\hat{C}_{ij},\hat{C}_{kl}]_\SN
    =\sum\limits_{
   i,j\in S_\epsilon\backslash I_1
    }
    \sum\limits_{
    k,l\in S_{\epsilon}\backslash I_2}
    \widehat{\llb{C}_{ij},{C}_{kl}\rrb}_\SN
   \nonumber\\
    &=\sum\limits_{
   i,j\in S_{\epsilon}\backslash I_1
    }
    \sum\limits_{
    k,l\in S_{\epsilon}\backslash I_2}
    (\delta_{j l}-\delta_{il})\sum\limits_{\scriptstyle \substack{ I\in\{x_i,y_i\}\\ J\in\{x_j,y_j\} 
    \\
     K\in\{x_k,y_k\}}}
     \hat{L}_{IJ}\odot  \hat{L}_{IK} \odot \hat{L}_{JK} 
     \nonumber\\
    &\quad+\sum\limits_{
   i,j\in S_{\epsilon}\backslash I_1
    }
    \sum\limits_{
    k,l\in S_{\epsilon}\backslash I_2} 
   (\delta_{j k}-\delta_{i k})
    \sum\limits_{\scriptstyle \substack{ I\in\{x_i,y_i\}\\ J\in\{x_j,y_j\} 
    \\
     L\in\{x_l,y_l\}}}
     \hat{L}_{IJ} \odot \hat{L}_{IL}\odot \hat{L}_{JL}
     \nonumber\\
     &=2\sum\limits_{
   i,j\in S_{\epsilon}\backslash I_1
    }
    \sum\limits_{
    k,l\in S_{\epsilon}\backslash I_2}
    (\delta_{j l}-\delta_{il})\sum\limits_{\scriptstyle \substack{ I\in\{x_i,y_i\}\\ J\in\{x_j,y_j\} 
    \\
     K\in\{x_k,y_k\}}}
     \hat{L}_{IJ}\odot  \hat{L}_{IK} \odot \hat{L}_{JK} 
     \nonumber\\
     &=4\!\!\!\sum\limits_{\substack{
   i\in I^c_1\backslash I^c_2\\
   j\in I^c_1\cap I^c_2\\
   k\in I^c_2\backslash I_1^c
   }}
   \sum\limits_{\scriptstyle \substack{ I\in\{x_i,y_i\}\\ J\in\{x_j,y_j\} 
    \\
     K\in\{x_k,y_k\}}}
     \hat{L}_{IJ}\odot  \hat{L}_{IK} \odot \hat{L}_{JK} 
    \,.
\end{align}
Here, we have introduced the notation $S_\epsilon=\{1,\dots,m+\epsilon\}$, $I^c_i:=S_\epsilon\backslash I_i$, and to change the index range, we have used the fact that if $i=j$ or $k=l$ then, as $C_{ii}$ or $C_{kk}$ commute with all other $C$'s, their commutator is zero. 
So we can add this zero to the sum to get the second equality. 
To show the final equality, one needs to expand the sum and observe that the double sums over the Kronecker deltas will not only eliminate one sum but also restrict the range of the remaining one to the intersection of the sets. 
Finally one notices that the antisymmetry of the terms in the summation on any exchange of two indices will reduce the sum to be over disjoint ranges.

We also mention this can be written explicitly in terms of the $\mathfrak{u}(m)$ generators:
\begin{equation}\label{eq: xi rho KT Comm LT }
     [K^{(I_1)}, K^{(I_2)}]_\SN   =4\!\!\!\sum\limits_{\substack{
   i\in I^c_1\backslash I^c_2\\
   j\in I^c_1\cap I^c_2\\
   k\in I^c_2\backslash I_1^c
   }}
   \big( 
   \hat{\rho}_{ij}\odot\hat{\rho}_{ik}\odot\hat{\xi}_{jk}
   -\hat{\rho}_{ij}\odot\hat{\rho}_{jk}\odot\hat{\xi}_{ik}
   +\hat{\xi}_{ij}\odot\hat{\xi}_{ik}\odot\hat{\xi}_{jk}
   +\hat{\xi}_{ij}\odot\hat{\rho}_{ik}\odot\hat{\rho}_{jk}
   \big)\,.
\end{equation}

Now, to complete the proof of \eqref{eq: KT Comms LT} we note that the final result in \eqref{eq: KT Comm LT} or \eqref{eq: xi rho KT Comm LT } is a sum over all three disjoint sets and so none of the terms can ever cancel. 
Therefore it is never zero unless the sum is over the empty set.
From set algebra, we have $I_i^c\backslash I_j^c\iff I_j\backslash I_i$ and therefore $I_1^c\backslash I_2^c$ (resp. $I_2^c\backslash I_1^c$) is empty if $I_2\subset I_1$ (resp. $I_1\subset I_2$). Thus we have proven the equivalence:   $[K^{(I_1)},K^{(I_2)}]=0$ iff $I_2\subset I_1$ or $I_1\subset I_2$.

Finally, we have an obvious collection of $m+\epsilon-1$ mutually commuting Killing tensors given by $I_i\in\left\{\emptyset,\{1\},\{1,2\},\dots,\{1,2,\dots, m+\epsilon-2\}\right\}$ or any other such collection of $m+\epsilon$ nested sets.
Physically speaking we can understand this as follows. 
The choice of subsets is a choice of which rotation axes to hold fixed: i.e. $\partial_{\phi_j}$ for $j\in I_i$. 
Then $K^{(I_i)}$ becomes the Casimir of the subgroup of rotations which holds fixed these same axes of rotation $\partial_{\phi_j}$. 
Thus, if each successive subset only fixes an additional axis, i.e. $I_{i+1}=I_i\cup \{k\}$ for some $\partial_{\phi_k}$, then the Casimirs will commute with each other. 
The Casimirs $K^{(I_i)}$ thus represent the totally angular momentum about the axis of rotation $\partial_{\phi_j}\in I_i$.

{The construction of the higher rank Killing tensors from the iterated SN brackets, as in \eqref{eq: higher KTs}, applies straightforwardly to this case. 
This is the origin of the growing tower of Killing tensors in the Lense--Thirring spacetimes.}

\subsection{Exact on-shell realization}\label{sec: on-shell}

The generalized Lense--Thirring spacetimes discussed above represent an off-shell example of geometry for which our construction works. 
It is well known, e.g. \cite{Gray:2021roq}, that upon choosing particular metric functions these embody approximate slowly rotating black hole solutions in various theories, see e.g. \eqref{Einstein} for slowly rotating black solutions in Einstein gravity. 
However, it is interesting to ask  whether any {\em exact} solution of possibly modified Einstein gravity with some matter fields can be cast in this form.  
Fortunately, at least in four dimensions, this is the case. In what follows we present an exact rotating black hole solution of the EMDA theory which takes the form of generalized Lense--Thirring spacetimes and for which our construction straightforwardly applies.

The EMDA theory is a scalar-vector-tensor theory of gravity, which arises from the low energy effective action of superstring theories, upon compactifying the ten-dimensional heterotic string theory on a six-dimensional torus \cite{Sen:1992ua}. 
It is described by the following action:
\begin{eqnarray}
S=\int d^4\!x \sqrt{-g}\Big[R-2\partial_{\mu}\phi\partial^{\mu}\phi-\frac{1}{2}e^{4\phi}\, \partial_{\mu}\kappa\partial^{\mu}\kappa\nonumber+e^{-2\phi}F_{\mu\nu}F^{\mu\nu}
+\kappa F_{\mu\nu}{(*F)}^{\mu\nu}\Big]\,,
\end{eqnarray}
where $F_{\mu\nu}$ is the field strength of $U(1)$ Maxwell field, and $\kappa$ and $\phi$ are the  axion and the dilaton, respectively. A generalization of the vacuum Kerr solution in this theory has been found by Sen \cite{Sen:1992ua}, while \cite{Galtsov:1994pd} represents the most general, locally asymptotically flat, extension known (see also \cite{Garcia:1995qz} and more recently \cite{Galtsov:2025nia}). Here, we are interested in a different exact rotating black hole solution in this theory, namely the one obtained in \cite{Clement:2002mb}.
It reads
\begin{eqnarray}
\label{EMDA}
&&g=-\frac{f}{r_0r}dt^2+r_0r\Big[\frac{dr^2}{f}+d\theta^2+\sin^2\!\theta\big(d\varphi-\frac{a}{r_0r}dt\big)^2\Big]\,,\nonumber\\
&& A=\frac{\sqrt{2}}{2}\Big(\frac{\rho^2}{r_0r}dt+a\sin^2\theta d\varphi\Big)\,,\nonumber\\
&& e^{-2\phi}=\frac{r_0r}{\rho^2}\,,\quad \kappa=-\frac{r_0 a \cos\theta}{\rho^2}\,,
\end{eqnarray}
where $\rho^2=r^2+a^2\cos^2\!\theta$ and $f=r^2-2mr+a^2$. 
This solution represents a rotating black hole with mass $M=\frac{m}{2}$ and angular momentum $J=\frac{a r_0}{2}$  \cite{Clement:2002mb}. 

Interestingly, upon the following coordinate transformation 
\be 
r_0r\to r^2\,,
\ee 
the metric takes the form
\ba 
g&=&-N\tilde fdt^2+\frac{dr^2}{\tilde f}+r^2\sin^2\!\theta\bigl(d\varphi-\frac{a}{r^2}dt\bigr)^2+r^2 d\theta^2\,,\nonumber\\
&&\tilde f=\frac{1}{4}-\frac{m r_0}{2r^2}+\frac{a^2r_0^2}{4r^4}\,,\quad N=\frac{4r^2}{r_0^2}\,,
\ea
which is precisely of the generalized Lense--Thirring form. 
Thus, the above EMDA solution represents an exact on-shell example of our construction, with the irreducible Killing tensor given by \cite{Clement:2002mb}:
\be 
K=\frac{1}{\sin^2\!\theta}(\partial_\varphi)^2+(\partial_\theta)^2=L_x^2+L_y^2+L_z^2\,,
\ee 
which originates from the underlying Killing vectors of the spherical transverse space $\cal S$; $L_z = \partial_\varphi, \;
L_x = \cot \theta \cos\varphi \partial_\varphi+\sin\varphi \partial_\theta,\; L_y = - \cot \theta \sin \varphi \partial_\varphi + \cos \varphi \partial_\theta$.

\section{More  examples }\label{Sec4}

We now turn to a discussion of other metrics which are of the form \eqref{Generic} starting with a physical example where the Killing vector symmetries of the base space lift to the full spacetime.
We then consider some illustrative but purely mathematical examples of applications of this formalism.

\subsection{Kerr-(A)dS metrics with equal rotation parameters}

It is well known that the symmetry of the Kerr-(A)dS metrics in $d=2m+1+\epsilon$ dimensions is enhanced from a $U(1)^m$ to a $U(m)$ symmetry~\cite{Vasudevan:2004ca,Galajinsky:2013mla}.
We point out here that this fits within our metric ansatz \eqref{Generic}. 
In particular, the Kerr--Ad(S) metric is given by~\cite{Gibbons:2004js,Gibbons:2004uw}:
\ba\label{eq: Kerr-AdS}
ds^2&=&-W(1+r^2/\ell^2)dt^2+\frac{2M}{U}\Bigl(Wdt+\sum_{i=1}^m \frac{a_i \mu_i^2 d\phi_i}{\Xi_i}\Bigr)^2\nonumber\\
&&+\sum_{i=1}^m\frac{r^2+a_i^2}{\Xi_i}\bigl(\mu_i^2d \phi_i^2+d\mu_i^2)
+\frac{Udr^2}{V-2M}+\epsilon r^2 d \mu_{m+\epsilon}^2\nonumber\\
&&-\frac{1}{W(\ell^2+r^2)}\Bigl(\sum_{i=1}^m \frac{r^2+a_i^2}{\Xi_i}\mu_i d\mu_i+\epsilon r^2 \mu_{m+\epsilon} d\mu_{m+\epsilon}\Bigr)^2\,, 
\ea 
where  $\ell$ is the AdS radius and
\ba
 W&=&\sum_{i=1}^m \frac{\mu_i^2}{\Xi_i}+\epsilon \mu_{m+\epsilon}^2\,,\quad V=r^{\epsilon-2}(1+r^2/\ell^2)\prod_{i=1}^m(r^2+a_i^2)\,,\nonumber\\
U&=&r^\epsilon\Bigl(\sum_{i=1}^{m+\epsilon}\frac{\mu_i^2}{r^2+a_i^2}\Bigr)\prod_{i=1}^{m}(r^2+a_i^2)\,,\quad \Xi_i=1-\frac{a_i^2}{\ell^2}\,. 
\ea
Here, as before, $\epsilon=1, 0$ for even, odd dimensions, $m=\bigl[\frac{d-1}{2}\bigr]$ , and the coordinates $\mu_i$ obey the same constraint as before \eqref{constraint}. 
Finally in these expressions $a_{m+1}=0=\phi_{m+1}$.

Next, we can use calculations similar to those detailed in Section 3.8 and Appendix A of \cite{Chrusciel:2025zfv} to put \eqref{eq: Kerr-AdS} into a form like \eqref{Generic}, namely:
\ba\label{eq: Kerr-AdS ADM}
ds^2&=&-f(r,\mu_i) N(r,\mu_i) dt^2+ \frac{dr^2}{f(r,\mu_i)} 
\nonumber
\\
&&+\sum_{i,j=1}^{m}h_{ij}(d\phi_i+\nu^i dt)(d\phi_j+\nu^j dt) +\sum_{i=1}^m\frac{r^2+a_i^2}{\Xi_i}d\mu_i^2+ \epsilon r^2 d \mu_{m+\epsilon}^2
\nonumber\\
&&-\frac{1}{W(\ell^2+r^2)}\Bigl(\sum_{i=1}^m \frac{r^2+a_i^2}{\Xi_i}\mu_i d\mu_i+\epsilon r^2 \mu_{m+\epsilon} d\mu_{m+\epsilon}\Bigr)^2\,.
\ea
Here we have introduced
\begin{align}
 f(r,\mu_i)&=\frac{V-2M}{U}\,,\quad
 N(r,\mu_i)=\frac{W\sum_{i=1}^{m+\epsilon}\frac{r^2\mu_i^2}{r^2+a_i^2}}{1+\frac{2M}{U}\sum_{i=1}^m\frac{a_i^2\mu_i^2}{(r^2+a_i^2)\Xi_i} }\,,
\end{align}
and
\begin{align}
h_{ij} d\phi_i d\phi_j 
	&= \sum_{i=1}^m
	\frac{r^2+a_i^2}{\Xi_i} 
	\mu_i^2 d\phi_i^2
    + \frac{2M}{U}
	\left( \sum_{i=1}^m \frac{a_i \mu_i^2 d\phi_i}{\Xi_i}   
	\right)^2 \,, 
    \nonumber\\
    \nu^i\partial_{\phi_i}&=\frac{2M}{U} \frac{N(r,\mu_i)}{\sum_{i=1}^{m+\epsilon}\frac{r^2\mu_i^2}{r^2+a_i^2}}\frac{a_i}{r^2+a_i^2}\partial_{\phi_i}\,. 
\end{align}

Some comments are in order. 
Although, the form of \eqref{eq: Kerr-AdS ADM} is similar to the generic form \eqref{Generic}, it is much messier. 
In particular, $f(r,\mu_i)$ and $N(r,\mu_i)$ depend on the transverse coordinates $\mu_i$, and the transverse metric $\gamma_{AB}$ itself ($\gamma_{AB}$ being  $h_{ij}$ and all the remaining terms in the second and third lines of \eqref{eq: Kerr-AdS ADM}) has no obvious symmetries to consider trying to lift to the full spacetime. 
Thus, although it is possible that the Killing tensor symmetries of Kerr lift from Killing vector symmetries of the transverse space we consider it unlikely, at least from this construction.

However, the situation is very different when all rotation parameters are equal, i.e. $a_i=a$.
The constraint implies that the terms in the bracket of the final line of \eqref{eq: Kerr-AdS ADM} are zero in odd dimensions because they are its derivative, i.e. $0=d(\sum_{i=1}^{m+\epsilon}\mu_i^2)=2\sum_{i=1}^{m+\epsilon}\mu_i d\mu_i$.
In even dimensions, they can be written together in terms of $\epsilon \mu_{m+\epsilon}^2d\mu_{m+\epsilon}^2$.
That is,
\begin{align}
    \frac{1}{W(\ell^2+r^2)}\Bigl(\sum_{i=1}^m \frac{r^2+a_i^2}{\Xi_i}\mu_i d\mu_i+\epsilon r^2 \mu_{m+\epsilon} d\mu_{m+\epsilon}\Bigr)^2
    =\frac{(r^2+\ell^2)}{(1-a^2/\ell^2)} \frac{\epsilon a^4\mu_{m+\epsilon}^2/\ell^2 }{(1-\epsilon a^2 /\ell^2\mu_{m+\epsilon}^2)} d \mu_{m+\epsilon}^2
\end{align}
Thus combining the $d\mu_i^2$ terms gives,
\begin{align}
    &\frac{r^2+a^2}{1-a^2/\ell^2}\sum_{i=1}^m d\mu_i^2 +\epsilon r^2 d\mu_{m+\epsilon}^2
    -\frac{(r^2+\ell^2)}{(1-a^2/\ell^2)} \frac{\epsilon a^4\mu_{m+\epsilon}^2/\ell^2 }{(1-\epsilon a^2 /\ell^2\mu_{m+\epsilon}^2)} d \mu_{m+\epsilon}^2 d \mu_{m+\epsilon}^2
    \nonumber\\
   =&\frac{r^2+a^2}{1-a^2/\ell^2}\sum_{i=1}^{m+\epsilon} d\mu_i^2 
   +\epsilon\left(-\frac{r^2+a^2}{1-a^2/\ell^2} + r^2 - \frac{(r^2+\ell^2)}{(1-a^2/\ell^2)} \frac{\epsilon a^4\mu_{m+\epsilon}^2/\ell^2 }{(1-\epsilon a^2 /\ell^2\mu_{m+\epsilon}^2)} \right)d \mu_{m+\epsilon}^2
   \nonumber\\
   =&\frac{r^2+a^2}{1-a^2/\ell^2}\sum_{i=1}^{m+\epsilon} d\mu_i^2
   -\epsilon \frac{a^2}{\ell^2}\frac{(r^2+\ell^2)}{(1-a^2/\ell^2)(1-\epsilon a^2\mu_{m+\epsilon}^2/\ell^2)}d\mu_{m+\epsilon}^2\,.
\end{align}

Hence, {in this case} we have 
\ba\label{eq: Kerr-AdS equal a}
ds^2&=&-f(r,\epsilon\mu_{m+\epsilon}^2) N(r,\epsilon\mu_{m+\epsilon}^2) dt^2+ \frac{dr^2}{f(r,\epsilon\mu_{m+\epsilon}^2)} 
\nonumber\\
&&+\sum_{i,j=1}^{m}h_{ij}(d\phi_i+\nu^i dt)(d\phi_j+\nu^j dt) +\frac{r^2+a^2}{\Xi}\sum_{i=1}^{m+\epsilon}d\mu_i^2 - \epsilon\Sigma(r,\epsilon\mu_{m+\epsilon}^2)\, d \mu_{m+\epsilon}^2\,,
\nonumber\\
\ea 
where now
\begin{align}
\Sigma(r,\epsilon\mu_{m+\epsilon}^2)&=\frac{a^2}{\ell^2}\frac{(r^2+\ell^2)}{(1-a^2/\ell^2)(1-\epsilon a^2\mu_{m+\epsilon}^2/\ell^2)}\,,
\end{align}
and
\begin{align}
W&=\frac{1-\epsilon a^2 \mu_{m+\epsilon}^2/\ell^2}{\Xi}\,,\quad  
\Xi=1-\frac{a^2}{\ell^2}\,, \quad U=r^{\epsilon-2}(r^2+a^2)^{m-1}(r^2+\epsilon a^2 \mu_{m+\epsilon}^2)
\,.
\end{align}
Notice that $N$ and $f$ are now functions of $r$ and $\epsilon\mu_{m+\epsilon}^2=\epsilon(1-\sum_i^{m+\epsilon-1}\mu_i^2)$, i.e. $N=N(r,\epsilon\mu_{m+\epsilon}^2)$,  $f=f(r,\epsilon\mu_{m+\epsilon}^2)$, whose explicit expressions are not important.
Moreover, this yields
\begin{align}
h_{ij} d\phi_i d\phi_j 
	&=\frac{r^2+a^2}{\Xi}  \sum_{i=1}^m \mu_i^2 d\phi_i^2
    + \frac{2M a^2r^{2-\epsilon}}{(r^2+a^2)^{m-1}(r^2+\epsilon a^2 \mu_{m+\epsilon}^2)\Xi^2}
	\left( \sum_{i=1}^{m} \mu_i^2 d\phi_i
	\right)^2 \,, 
    \nonumber\\
    \nu^i\partial_{\phi_i}&= \frac{2M N(r,\epsilon\mu_{m+\epsilon}^2)}{r^{\epsilon}(r^2+a^2)^{m-1}} \partial_{\phi_i}\,.
\end{align}
{One can check that this recovers the expressions in \cite{Galajinsky:2013mla} in the asymptotically flat limit, i.e. when $\ell\to\infty$.}

Now, in odd dimensions it is clear that the metric is in the form of \eqref{Generic} since $\epsilon=0$ the dependence on $\mu_i$ in $N$, $f$, and $\nu^i$ is gone.%
\footnote{In odd dimensions the metric has also be rewritten into the dual ansatz form presented in Appendix \ref{app: dual met} \eqref{Generic dual}, see e.g. \cite{Kunduri:2006qa}.
}
In fact, we shall show that certain symmetries of the transverse metric lift to symmetries of $\cal M$ in both even and odd dimensions.

The question is then: what are the Killing vectors of the transverse metric? To answer this, we observe the transverse metric of \eqref{eq: Kerr-AdS equal a} is
\begin{align}\label{eq: trans met Kerr-AdS equal a}
\gamma_{AB} dx^A dx^B=&\underbrace{\frac{r^2+a^2}{1-\frac{a^2}{\ell^2}}\left(\sum_{i=1}^{m+\epsilon}d\mu_i^2+\sum_{i=1}^m\mu_i^2d\phi_i^2\right)-{\epsilon \Sigma(r,\epsilon\mu_{m+\epsilon}^2)\, d \mu_{m+\epsilon}^2}}_{:=\mathring{\gamma}_{AB} dx^Adx^B}
\nonumber\\
& + \frac{2M a^2r^{2-\epsilon}}{(r^2+a^2)^{m-1}(r^2+\epsilon a^2 \mu_{m+\epsilon}^2)\Xi^2}
	\left( \sum_{i=1}^m \mu_i^2 d\phi_i
	\right)^2\,.
\end{align}

The first line of this equation is, in odd dimensions, simply the metric on the $m$ sphere $\mathring{\gamma}_{AB}$. 
Thus it has the same Killing vectors \eqref{eq: L KVs} for $i,j\leq m$. The extra $\epsilon d\mu_{m+\epsilon}^2$ in even dimensions breaks this symmetry. 
However, one can easily check, using Cartan's formula for the Lie derivative, that the sub-algebra $\mathfrak{u}(m)$ generators  (\eqref{eq: xi KV} and \eqref{eq: rho KV}) are still symmetries. 
Clearly they are symmetries of the bracketed term in both even and odd dimensions and moreover,
\begin{align}\label{eq: xi rho mu}
    {\cal L}_{\xi}(d\mu_{m+\epsilon}^2)&=i_\xi d^2(\mu_{m+\epsilon}^2)+d(i_\xi(\mu_{m+\epsilon}^2))=0\,,
    \nonumber\\
    {\cal L}_{\rho}(d\mu_{m+\epsilon}^2)&=i_\rho d^2(\mu_{m+\epsilon}^2)+d(i_\rho(\mu_{m+\epsilon}^2))=0\,.
\end{align}
The last equality in each case follows from using the explicit expressions of $\xi$ and $\rho$  (\eqref{eq: xi KV} and \eqref{eq: rho KV}) to see that
\be
\xi^A\partial_A(\mu_{m+\epsilon}^2)=0=\rho^A\partial_A(\mu_{m+\epsilon}^2)\,.
\ee
{This also implies the Lie derivative along $\xi$ or $\rho$ of any function of $\mu_{m+\epsilon}^2$ is zero, e.g. $\Sigma(r,\epsilon\mu_{m+\epsilon}^2)$. Hence, their Lie derivatives of the final term in $\mathring{\gamma}$ vanishes.}
Thus,  $\xi$ and $\rho$ are Killing vectors of $\mathring{\gamma}$.

Next, one can see that the off-diagonal $d\phi_id\phi_j$ terms in \eqref{eq: trans met Kerr-AdS equal a} are of the form
\be
h(r,\epsilon \mu_{m+\epsilon}^2 )  (\mathring{\gamma}_{AB}\zeta^B dx^A)^2\,,\quad \text{where } \zeta^A\partial_A=\sum_{i=1}^m\partial_{\phi_i}=\frac{1}{2}\sum_{i=1}^{m}\rho_{ii}\,,
\ee
for some function $h(r,\epsilon\mu_{m+\epsilon}^2)$ whose particular form is unimportant.
Then one can immediately check, using \eqref{eq: u(m) algebra} that the $\mathfrak{u}(m)$ generators  ($\xi$ and $\rho$) commute with $\zeta$. 
That is,
\begin{align}
    \llb \xi_{ij},\zeta\rrb&=\frac{1}{2}\sum_{k=1}^m\llb\xi_{ij},\rho_{kk}\rrb=\sum_{k=1}^m(\delta_{jk}\rho_{ik}-\delta_{ik}\rho_{jk})=\rho_{ji}-\rho_{ij}=0\,,
    \nonumber
    \\
     \llb \rho_{ij},\zeta\rrb&=\frac{1}{2}\sum_{k=1}^m\llb\rho_{ij},\rho_{kk}\rrb=-\sum_{k=1}^m(\delta_{jk}\xi_{ik}+\delta_{ik}\xi_{jk})=-(\xi_{ji}+\xi_{ij})=0\,.
\end{align}
As mentioned above, they also Lie commute with any function $h(\epsilon \mu_{m+\epsilon}^2)$ and hence are Killing vectors not only of $\mathring{\gamma}$ but also of the full $\gamma_{AB}$ in \eqref{eq: trans met Kerr-AdS equal a}.%
\footnote{
{In four dimensions these results apply to Kerr-(A)dS, in particular \eqref{eq: trans met Kerr-AdS equal a} remains true. However, the only non-zero vector is the $U(1)$ symmetry of $\partial_\phi$. So the only obvious symmetry of the base space is the usual axi-symmetry.}
}

Furthermore, this means that $\xi$ and $\rho$ lift to Killing vectors of the full spacetime, since $\nu=\tilde{h}(r,\epsilon\mu_{m+\epsilon}^2)\zeta$ and so for precisely the same reasons 
\begin{align}
&\llb \xi,\nu\rrb=0=\llb \rho, \nu \rrb\,, \quad \xi^A\partial_A f(r,\mu_{m+\epsilon}^2)=0=\rho^A\partial_Af(r,\mu_{m+\epsilon}^2)\,,
\nonumber\\
& \xi^A\partial_A N(r,\mu_{m+\epsilon}^2)=0=\rho^A\partial_AN(r,\mu_{m+\epsilon}^2)\,.
\end{align}
and thence making use of the decomposition \eqref{eq: Lifted bracket}, we see
\begin{equation}
   [ \hat{\xi},g^{-1}]_{\SN}=0=[ \hat{\rho},g^{-1}]_{\SN}\,.
\end{equation}

Therefore, we have shown that the $U(m)$ symmetry of Myers--Perry black holes with equal rotation parameters can be understood as the lifting of the $U(m)$ symmetry of the transverse metric to the full spacetime. At the same time, the corresponding Casimirs give rise to reducible Killing tensors of the full spacetime.
We expect that similar conclusions can also be derived in less symmetric cases, such as those when there are two sets of equal rotation parameters \cite{Vasudevan:2004mr}, in which case, however, there already exists a single irreducible Killing tensor.

\subsection{Planar symmetry}
\subsubsection*{Four-dimensional example}
To discuss our first `purely mathematical example',  we consider the following metric:
\be\label{eq:planar}
ds^2 = g_{\mu\nu} dx^\mu dx^\nu = -Nf dt^2 + \frac{dr^2}{f} + r^2 d\rho^2 + r^2 \rho^2 (d\phi + \omega dt)^2\,,
\ee
where $\omega = \omega(r)$ is a function of the radial coordinate. 

In the special case where $\omega = 0$, the metric exhibits four Killing vector fields, corresponding to the isometries $\partial_t$, $\partial_\phi$, $\partial_x$, and $\partial_y$, where we introduce Cartesian coordinates $x = \rho \cos\phi$ and $y = \rho \sin\phi$. 
These symmetries reflect the underlying planar invariance of the two-dimensional spatial slices parametrised by $(x,y)$. 
Specifically, $\partial_x$ and $\partial_y$ generate translations in the Cartesian plane, while $\partial_\phi$ corresponds to rotational symmetry about the origin, and $\partial_t$ represents time translation invariance.

However, when $\omega \neq 0$, the presence of  additional terms in the metric explicitly breaks the full planar symmetry. 
In particular, the full Killing algebra no longer includes $\partial_x$ and $\partial_y$ as global symmetries of the spacetime. 
Nevertheless, the two-dimensional spatial metric, given by
\be 
\gamma=\gamma_{AB} dx^A dx^B = r^2 (d\rho^2 + \rho^2 d\phi^2) = r^2 (dx^2 + dy^2)\,,
\ee
remains invariant under translations in the $(x,y)$ plane. 

Using the notation from Sec. \ref{sec: Generic Metrics}, the shift vector in \eqref{eq:planar} is given by $\nu=\omega(r)\partial_\phi$. 
Killing vector $\partial_\phi$ does not commute with $\partial_x$ and $\partial_y$ separately, but it does commute with their quadratic combination given by $C\equiv\partial_x^2+\partial_y^2=r^2 \gamma^{-1}$:
\begin{align}
    &\llb \partial_\phi,\partial_x\rrb=-\partial_y, \quad \llb \partial_\phi,\partial_y\rrb=\partial_x\,,\nonumber\\
    &\llb\partial_\phi,C \rrb_{\SN}= \llb\partial_\phi,\partial_x^2 \rrb_{\SN}+\llb\partial_\phi,\partial_x^2 \rrb_{\SN}=-2\partial_x\partial_y+2\partial_x\partial_y=0\,,
\end{align}
with the latter obvious, as $\partial_\phi$ is a Killing vector of $\gamma$.
Since it is a linear combination of Killing vectors, $C$ is a rank-two reducible Killing tensor of the 2-planes. 
Its commutativity with $\partial_\phi$ ensures that this rank-two reducible Killing tensor of the 2-planes lifts to an irreducible Killing tensor of the full spacetime 
\be\label{eq: K planar}
\hat C=\partial_x^2+\partial_y^2\,.
\ee

The implications for geodesic motion are as follows. 
The 2-dimensional base space admits 3 independent integrals of motion, associated with for example the following set of independent symmetries: $\{\partial_\phi, \partial_x, C\}$. 
Of these one can choose 2 commuting: $\{\partial_\phi, C\}$. 
As expected, the motion on ${\cal S}$ is thus maximally superintegrable. 
Upon the lift to ${\cal M}$ we now have the following set of 4 independent and mutually commuting symmetries: $\{\partial_t, \partial_\phi, \hat C, g^{-1}\}$, while $\partial_x$ is no longer a Killing vector. 
The motion in ${\cal M}$ is thus `only' completely integrable.

The spacetime \eqref{eq:planar} was identified in \cite{He:2024tab} in the context of broken planar symmetry, and shown to admit a hidden symmetry encoded in the rank-two Killing tensor matching our expression for $\hat C$. 
Here we have shown that this hidden symmetry stems from explicit symmetries in the two-dimensional leaves.%
\footnote{Moreover, one can check that the generalization of \eqref{eq:planar} introduced in \cite{He:2024tab}:
\be 
ds^2 = -Nf dt^2 + \frac{dr^2}{f} + r^2 d\rho^2 + \frac{r^2\rho^2}{1+s(r)\rho^2} (d\phi + \omega dt)^2\,,
\ee
represents the most general deformation of the transverse metric, $\gamma_{AB}dx^Adx^B=r^2(d\rho^2+\rho^2/(1+s(r)\rho^2)d\phi^2$, which breaks the Killing vector symmetry of $\partial_x$ and $\partial_y$,  such that $C$ in \eqref{eq: K planar} remains a Killing tensor. The lifting of $C$ 
to a Killing tensor of the full metric $\hat C$ is immediate in our construction. 
}

\subsubsection*{Higher dimensions}
The previous example generalizes naturally to higher dimensions. 
The metric now takes the form:
\be \label{eq:nplanar}
ds^2 = -N f dt^2 + \frac{dr^2}{f} + r^2 d\Sigma^2_{(n)} + r^2 \rho^2 (d\theta_{n-1} + \omega dt)^2\,, 
\ee
where $ d\Sigma^2_{(n)} $ represents the metric of an $n$-dimensional Euclidean space in polar coordinates, given by  
\be 
d\Sigma^2_{(n)} = d\rho^2 + \rho^2 d\Omega_{n-1}^2\,. 
\ee  
Here, $ d\Omega_{n-1}^2 $ is the standard metric on the $(n-1)$-dimensional unit sphere, explicitly written as  
\be 
d\Omega_{(n-1)}^2 = d\theta_1^2 + \sin^2\theta_1 d\theta_2^2 + \sin^2\theta_1 \sin^2\theta_2 d\theta_3^2 + \dots + \prod_{i=1}^{n-2} \sin^2\theta_i d\theta_{n-1}^2\,. 
\ee

The shift vector in \eqref{eq:nplanar} is again proportional to a Killing vector,  
\be
\nu = \omega(r) \partial_{\theta_{n-1}}\,.
\ee  
To analyse the isometries, we introduce Cartesian coordinates $(x_1,\dots,x_n)$ related to the spherical coordinates by  
\begin{align} 
x_1 &= \rho \cos\theta_1\,,\nn\\  
x_2 &= \rho \sin\theta_1 \cos\theta_2\,, \nn\\
&\vdots \nn\\
x_{n-1} &= \rho \sin\theta_1 \sin\theta_2 \cdots \sin\theta_{n-2} \cos\theta_{n-1}\,, \\
x_n &= \rho \sin\theta_1 \sin\theta_2 \cdots \sin\theta_{n-2} \sin\theta_{n-1}\,.\nn
\end{align}

Isometries of the $n$-dimensional Euclidean space correspond to the generators $ \partial_{x_i} $. The special Killing vector $ \partial_{\theta_{n-1}} $ induces a rotation in the $(x_{n-1}, x_n)$-plane, leading to the following commutation relations:  
\be 
\llbracket \partial_{\theta_{n-1}}, \partial_{x_{n-1}} \rrbracket = -\partial_{x_n}\,, \quad 
\llbracket \partial_{\theta_{n-1}}, \partial_{x_n} \rrbracket = \partial_{x_{n-1}}\,, \quad 
\llbracket \partial_{\theta_{n-1}}, \partial_{x_i} \rrbracket = 0 \quad\text{for } i < n-1\,. 
\ee
Vectors $ \partial_{x_i} $ for $ i < n-1 $ remain Killing vectors of the full spacetime due to their commutativity with the shift vector. 
However, $ \partial_{x_{n-1}} $ and $ \partial_{x_n} $ do not, as they mix under the action of $ \partial_{\theta_{n-1}} $. 
Nevertheless, their quadratic combination remains a symmetry and lifts to a rank-two Killing tensor. Specifically, defining  
$C \equiv \partial_{x_{n-1}}^2 + \partial_{x_n}^2$,  
we find that it remains invariant under the action of $ \partial_{\theta_{n-1}} $:  
\be
\llbracket \partial_{\theta_{n-1}}, C \rrbracket_{\SN} = \llbracket \partial_{\theta_{n-1}}\,, \partial_{x_{n-1}}^2 \rrbracket_\SN + \llbracket \partial_{\theta_{n-1}}, \partial_{x_n}^2 \rrbracket_\SN = -2\partial_{x_{n-1}} \partial_{x_n} + 2\partial_{x_n} \partial_{x_{n-1}} = 0\,.
\ee

Again, $ C $ is a  rank-two reducible Killing tensor in the $n$-dimensional Euclidean space. 
Its commutativity with $ \partial_{\theta_{n-1}} $ ensures that it lifts to an irreducible Killing tensor of the full spacetime:  
\be
\hat{C} = \partial_{x_{n-1}}^2 + \partial_{x_n}^2.
\ee

\subsection{Lifting the Taub--NUT metric}
\label{subsec: Taub--NUT}
To emphasise the utility of the formalism, let us now construct a final mathematical example. 
A recent analysis of symmetry reductions of the gravitational Lagrangian \cite{Frausto:2024egp} offers a compelling starting point for selecting the symmetry group of the base space in our construction. Their classification of admissible isometry groups can be used as a guide for building spacetimes with hidden symmetries within our framework. 
While their approach is formulated in Lorentzian signature, it can be adapted to our setting either by implementing a timelike foliation in our construction or by extending their methods to spacetimes with Euclidean signature. 
One of the viable symmetry groups identified there is that of Taub-NUT spacetime. 

Here we start from its Euclideanized version~\cite{Taub:1950ez,Newman:1963yy,Gibbons:1987sp} discussed in  Appendix~\ref{app: dual met}, where we set $n\to in$ and $t\to i\rho$. 
Let us also (for simplicity) suppress the radial direction of the full Taub--NUT, that is, we consider $x^A=\{\varphi, \theta, \rho\}$. 
Then we have 
\be 
\gamma=\gamma_{AB}dx^A dx^B=g(d\rho-2n \cos\theta d\varphi)^2+(r^2-n^2)(d\theta^2+\sin^2\theta d\varphi^2)\,.
\ee 
Here {we have introduced an arbitrary function} $g=g(t,r)$ and $t,r$ are treated as external parameters. 
The metric $\gamma$ admits 4 Killing vectors corresponding to $SU(2)\times \mathbb{R}$ isometry, namely:
\ba 
\xi_1 &=& - \sin \varphi \cot\theta\partial_\varphi + \cos \varphi \partial_\theta - 2n \sin \varphi\frac{
1 - \cos \theta}{\sin\theta}\partial_\rho\,,\nonumber\\
\xi_2 &=& \cos \varphi \cot \theta \partial_\varphi + \sin \varphi \partial_\theta + 2n \cos \varphi\frac{1-\cos \theta}{\sin\theta}\partial_\rho\,,\nonumber\\
\xi_3 &=& \partial_\varphi
- 2n\partial_\rho\,,\quad \eta=\partial_\rho\,,
\ea
see \cite{Jezierski:2006fw,Gibbons:1987sp,Kalamakis:2020aaj}
for a complete discussion of the symmetries of $\gamma_{AB}$.
We also have 
\be 
\llb C,\xi_3\rrb_\SN=0\,, \quad C=\sum_i \xi_i^2\,.
\ee 
Then, choosing $\nu=p(t,r)\xi_3$ the following spacetime:
\be 
ds^2=-Nf dt^2+\frac{dr^2}{f}+\gamma_{AB}\Bigl(dx^A+p [\delta^A_\varphi - 2n\delta^A_\rho] dt\Bigr)\Bigl(dx^B+p [\delta^B_\varphi - 2n\delta^B_\rho] dt\Bigr)\,,
\ee 
where $f=f(t,r), N=N(t,r)$, and $p=p(t,r)$ are arbitrary functions of $t$ and $r$, satisfies all conditions of our construction.
In particular it will admit a non-trivial Killing tensor, {given by $\hat C$}.
Of course one could generalize this {example to arbitrary dimensions using the $(d-2)$-dimensional version of the Euclideanized Taub--NUT, e.g. \cite{Awad:2000gg,Bueno:2018uoy} (suppressing potentially its radial direction which plays no role).

\section{Discussion and summary}
\label{sec: conclusion}

In this work we have investigated the emergence of irreducible Killing tensors from lower-dimensional Killing vectors.
While explicit spacetime symmetries are described by Killing vectors, many spacetimes exhibit hidden symmetries encoded in higher-rank Killing tensors. 
Recent studies on Lense–Thirring spacetimes have suggested the existence of an infinite hierarchy of such tensors, but their geometric origin had remained unclear.

In contrast to previous related ideas~\cite{Beig:1996ew,Chrusciel:2013aya,Kobialko:2021aqg,Bogush:2021qnz} we have proposed a systematic construction based on a spacetime ansatz foliated by codimension-2 hypersurfaces. 
Our key finding is that Killing tensors, which are reducible to products of non-commuting Killing vectors in the base space, naturally lift to irreducible Killing tensors of the full spacetime.
This structure is dictated by the Lie algebra of the base space symmetries, leading to a tower of higher-rank Killing tensors. 
We then applied this framework to generalized Lense–Thirring spacetimes and {demonstrated its realization in exact four-dimensional EMDA} black hole solutions. {It remains to be seen, if any other irreducible Killing tensors of exact black hole spacetimes can be understood in this way.}   
Moreover, the $U([d/2]-1)$ symmetry of Kerr-(A)dS spacetimes with equal angular momenta was shown to emerge from the same such symmetry of the {lower-dimensional hypersurface}.

These results provide a new geometric understanding of hidden symmetries in higher-dimensional spacetimes and have several natural extensions.
In particular, one could consider topological Lense--Thirring spacetimes, {where the cross-sectional metric $\gamma_{AB}$ on $\cal S$ has for example negative constant curvature.}
Such metrics solve, for example, the linearised Einstein equations (see appendix H of~\cite{Cong:2024xam}) and will inherit in exactly the same way the symmetries of $\gamma_{AB}$.
Of course, if $\cal S$ is compact and negatively curved it has no (global) Killing vectors.
We expect this would also lead to new topological slowly rotating black holes in many such theories beyond Einstein gravity (e.g, those discussed in \cite{Gray:2021roq}) with lifted symmetries.

More generally speaking, can one construct initial data \`a la \cite{Beig:1996ew} for higher order symmetries i.e. Killing tensors? And under what conditions would Killing--Yano tensors lift to the full spacetime for our metric ansatz \eqref{Generic}?
Furthermore, we could seek to adapt the geometric lifting procedure in \cite{Kobialko:2021aqg,Bogush:2021qnz} to the general co-dimension-2 foliations outlined in Appendix \ref{app: foliation}. 
It would be natural to expect that geometrically nice (e.g. geodesic) foliations lead to natural lifts of symmetries of the transverse space.
These would be applicable to such special coordinate systems like Gaussian null (Isenberg--Moncrief)~\cite{Moncrief:1983xua}  and Bondi~\cite{Bondi:1962px,Sachs:1962zza}, which have become ubiquitous for understanding important physical situations.

Finally, continuing this theme, we could seek to understand the emergence of near horizon/near future null infinity symmetries via our geometric construction. In these asymptotic regimes, the spacetime geometry simplifies and can be effectively described by a fibration over a transverse spatial surface $\cal S$. 
Using our construction one could study possible extensions of asymptotic symmetry algebras, such as the BMS group at null infinity~\cite{Bondi:1962px,Sachs:1962zza,Strominger:2017zoo}, to include higher-rank symmetry generators. 
It would be interesting to investigate whether lifted Killing tensors associated with the transverse space in codimension-2 foliations could give rise to a novel class of asymptotic symmetries, potentially forming a higher-spin or tensorial extension of the BMS algebra, similar to what was shown for flat space in \cite{Campoleoni:2021blr}, or generalizing the Killing tensor surface charges presented in ~\cite{Barnich:2005bn} for constant curvature spacetimes. Such a structure could lead to new insights into conserved quantities relevant for gravitational radiation, memory effects, and soft theorems.
We leave these directions for future work.

\section*{Acknowledgements}
We thank Atharva Patil for carefully reading the manuscript and identifying an error prior to publication.
G.O. is grateful for support from GAČR 23-07457S and P.K. thanks to GAČR 22-14791S grants of the Czech Science Foundation. {D.K. is grateful for support from GAČR 23-07457S grant of
the Czech Science Foundation and  the Charles University Research Center Grant No. UNCE24/SCI/016. 
D.K. would also like to thank the Perimeter Institute for Theoretical Physics for
hospitality, where part of this work was completed. Research at Perimeter Institute is supported by the Government of Canada through the Department of Innovation, Science and Economic Development and by the
Province of Ontario through the Ministry of Colleges and Universities.
}

\appendix

\section{Foliation of spacetime by codimension-2 hypersurfaces}\label{app: foliation}
\subsection{Construction}

In this appendix we follow the conventions set up in Chapter 2 of \cite{Odak:2023jgs}.
We consider a $(2 + (d-2))$ foliation of spacetime induced by two scalar fields $\psi^a$, with $a = 0,1$. These define two closed one-forms, $n^a = d\psi^a$, which we interpret as normal covectors to a family of codimension-2 surfaces, denoted $\cal S$.

The symmetric lapse matrix is defined by 
\begin{equation} 
N^{ab} \equiv n^a_\mu n^{b\mu}\,, 
\end{equation} and assumed to be non-degenerate with $\det N_{ab} < 0$, so that the surfaces $\cal S$ are spacelike. The inverse matrix $N^{ab}$ is used to define the dual basis of vector fields
\begin{equation} 
n_a^\mu \equiv N_{ab} g^{\mu\nu} n_\nu^b\,, 
\end{equation} 
where each $n_a^\mu$ is tangent to the hypersurface $\psi^b = \text{const}$ for $b \neq a$.
We introduce the projector onto the codimension-2 leaves $\cal S$ as 
\begin{equation} 
\gamma^\mu_\nu \equiv \delta^\mu_\nu - N_{ab} n^a_\nu n^{b\mu}, 
\end{equation} 
with the induced metric on $S$ given by 
\begin{equation} 
\gamma_{\mu\nu} \equiv \gamma^\lambda_\mu \gamma^\sigma_\nu g_{\lambda\sigma}. 
\end{equation}
The complement of the surfaces ${S}$ is a two-dimensional plane spanned by $n^\mu_a$, which need not be integrable. The obstruction to integrability is the twist, defined through the commutator 
\begin{equation} 
[n_0, n_1]^\mu \notin \text{Span }\{n^\mu_a\}, \quad \text{or equivalently} \quad \gamma^\mu_\nu [n_0, n_1]^\nu \neq 0. 
\end{equation}

To facilitate coordinate expressions, we introduce a set of coordinates adapted to the foliation, $(\psi^a, x^A)$, where $x^A$ parameterize the leaves $\cal S$, and $\psi^a$ serve as foliation parameters. The coordinate vectors $\partial_\psi^a$ need not be orthogonal to $\cal S$; this motivates the introduction of shift vectors
\begin{equation}
b^\mu_a \equiv (\partial_{\psi^a})^\mu - n^\mu_a. 
\end{equation}

In this coordinate system, the spacetime metric takes the form \begin{equation} 
g_{\mu\nu} = 
\begin{pmatrix} 
N_{ab} + \gamma_{AB} b^A_a b^B_b &\quad \gamma_{AC} b^C_b \\
\gamma_{BC} b^C_a & \gamma_{AB} 
\end{pmatrix}, \quad g^{\mu\nu} = 
\begin{pmatrix} 
N^{ab} & -N^{a\mu} b_\mu^B \\
-N^{b\mu} b_\mu^A &\quad \gamma^{AB} + N^{\mu\nu} b_\mu^A b_\nu^B 
\end{pmatrix}, 
\end{equation} 
with the determinant 
\begin{equation} 
g = \det N_{ab} \det \gamma_{AB}. 
\end{equation}
We also note the expression for the inverse relation 
\begin{equation} 
n^\mu_a = N_{ab} g^{\mu\nu} n_\nu^b = N_{ab} g^{\mu b}, 
\end{equation} 
and the expression for the commutator of the normal vectors 
\begin{equation} \label{eq:genint} 
[n_a, n_b]^\mu = -2\partial_{[a} b_{b]}^\mu + [b_a, b_b]^\mu, 
\end{equation} with $b^\mu_a = (0, b^A_a)$ in adapted coordinates.

\subsection{Gauge fixing}
Up to this point, the formalism is fully general and no gauge fixing has been imposed. To connect with the metric ansatz discussed in Section \ref{sec: Generic Metrics}, we now impose a specific choice of coordinates $(t, r, x^A)$ and gauge conditions:

\begin{enumerate}
    \item We fix the off-diagonal components of the lapse matrix $N_{tr}=N_{rt}=0.$
    \item We choose the radial shift vector to vanish $b_r^\mu=0$, which amounts to taking the coordinates $x^A$ constant along the radial direction.
\end{enumerate}
With this, we identify: 
\begin{align} 
b_t^A &= \nu^A, \\ 
N_{tt} &= -N f, \\ 
N_{rr} &= \frac{1}{f}. 
\end{align}
The resulting parametrized metric becomes 
\begingroup \renewcommand*{\arraystretch}{1.5} 
\begin{equation} 
g_{\mu\nu} = \begin{pmatrix} 
-Nf + r^2 \gamma_{CD} \nu^C \nu^D &\quad 0 &\quad  \gamma_{AC} \nu^C \\
0 &\quad \frac{1}{f} &\quad 0 \\
 \gamma_{BC} \nu^C &\quad 0 &\quad  \gamma_{AB} \end{pmatrix}, 
\end{equation}

\begin{equation} 
g^{\mu\nu} = \begin{pmatrix} 
-\frac{1}{Nf} &\quad 0 &\quad \frac{1}{Nf} \nu^B \\
0 &\quad f &\quad 0 \\
\frac{1}{Nf} \nu^A &\quad 0 &\quad  \gamma^{AB} - \frac{1}{Nf} \nu^A \nu^B \end{pmatrix}, 
\end{equation} 
\endgroup
which agrees with \eqref{Generic} and \eqref{Generic inverse} in the main text.
{Here, however for full generality the functions $f$ and $N$ depend on all the coordinates $x^\mu=(t,r,x^A)$.}

From the general integrability condition in Eq. \eqref{eq:genint}, we observe that the normal 2-planes are integrable if and only if $\nu^A$ is independent of the radial coordinate $r$. However, if $\nu^A$ has the form 
\begin{equation} 
\nu^A = p(t,r) \xi_0^A(x^B), 
\end{equation} 
as assumed in the main text, then the twist is non-vanishing and proportional to the Killing vector $\xi_0^A$. Explicitly, 
\begin{equation} 
[n_r, n_t]^\mu = -\partial_r \hat{\nu}^\mu = -\frac{\partial p(t, r)}{\partial r} \hat{\xi}_0^\mu. 
\end{equation} 

\section{Alternative construction}\label{app: dual met}

\subsection{Dual metric ansatz}

{In this appendix we briefly comment on a possible dual formulation of our general ansatz \eqref{Generic}. 
Namely, let us consider the following metric:
\be\label{Generic dual}
\tilde g=-\tilde{f} \tilde{N} (dt+\tnu_Adx^{A})^2+\frac{dr^2}{\tilde{f}}+\tgamma_{AB}dx^A dx^B\,,
\ee
c.f. the inverse metric \eqref{Generic inverse}. 
Here, we again assume $\tilde f$ and $\tilde N$ to be functions of $t, r$ only and ${\tilde \nu}^A$ and ${\gamma}_{AB}$ to be functions of all coordinates. 
Moreover, $\tnu_A$ is naturally a 1-form while previously $\nu^A$ was naturally a vector.
The inverse now reads
\be\label{Generic inverse dual}
{\tilde g}^{-1}=-\frac{1}{\tilde{N}\tilde{f}}\partial_t^2+\tf\partial_{r}^2+\tgamma^{AB}(\partial_A-\tnu_A \partial_t)(\partial_B+\tnu_B \partial_t)\,.
\ee
Notice that the structures are more or less the same, except by comparing the $\partial_t^2$, $\partial_t\partial_{A}$, $\partial_{A}\partial_{B}$ terms we have
\be
(\tf \tN)^{-1}-\tnu^A\tgamma_{AB}\tnu^B= (fN)^{-1}\,, \quad  \tnu^A=-(fN)^{-1}\nu^A\,,\quad \tgamma^{AB}-\tf\tN\tnu^A\tnu^B=\gamma^{AB}\,,
\ee
where $\gamma^{AB}=(\gamma^{-1})^{AB}$ and $\tgamma^{AB}=(\tgamma^{-1})^{AB}$.
Thus one can translate from one description to another. 
However, if the transverse metric is fixed (e.g. in both case setting it to be the metric on the sphere) then symmetries in one description will not be symmetries in the other unless $\tnu^2\equiv\tnu^A\tgamma_{AB}\tnu^B$ is $x^A$ independent.}

In this case the decomposition of the SN bracket \eqref{eq: Lifted bracket} becomes
\begin{align}
    [\hat{X}, g^{-1}]_{\SN}^{\mu_1\dots \mu_{p+1}}\partial_{(\mu_1\dots\mu_{p+1})}
    &-\llb X, (\tf\tN)^{-1} -\tnu^2\rrb^{ A_1\dots A_{p-1} }_{\SN}\,\, \partial_{(t}\partial_t\partial_{A_1}\dots\partial_{A_{p-1})}
    \nonumber\\
    &+\llb X, \tf\rrb^{ A_1\dots A_{p-1} }_{\SN}\,\, \partial_{(r}\partial_r\partial_{A_1}\dots\partial_{A_{p-1})} 
    \nonumber\\
    &+ 2\left((\tf\tN)^{-1}\partial_t X^{ A_1\dots A_{p}}-\llb X, \tnu \rrb^{ A_1\dots A_{p} }_{\SN} \right)\partial_{(t}\partial_{A_1}\dots\partial_{A_{p})}
    \nonumber\\
    &-2\tf \left(\partial_r X^{ A_1\dots A_{p}}\right) \,\, \partial_{(r}\partial_{A_1}\dots\partial_{A_{p})}
    \nonumber\\
    &+2\tnu^{(A_1}\partial_t X^{A_2\dots A_{p+1})} \partial_{(A_1}\dots\partial_{A_{p+1})}\,
    \nonumber\\
    &+\llb X, \tgamma^{-1} \rrb_\SN^{A_1\dots A_{p+1}}\partial_{(A_1}\dots\partial_{A_{p+1})}\,.
\end{align}
Hence with this dual choice of metric ansatz, the conditions for a symmetry of $\cal S$ to lift to the full spacetime $\cal M$ are \emph{almost} exactly the same as previously. Namely,
 $\llb X, \tf\tN \rrb_{\SN}=0$ (again automatically satisfied as
$\tf, \tN$ depend on $t,r$ only, i.e. $\tf=\tf(t,r)$ and $\tN=\tN(t,r)$), and
\begin{equation}\label{eq: dual Sym conditions 0}
  \partial_t X=0\,, \quad \partial_r X=0\,,\quad  \llb X, \tnu^2\rrb_\SN=0\,,\quad\llb X, \tnu \rrb_{\SN}=0\,.
\end{equation}

It is worth now discussing the third condition in more detail. 
If $\tnu^2$ is $x^A$ independent then it is trivial. 
But, in the case where $X=\sum_i\xi_i\otimes \xi_i$ for Killing vectors $\xi_i$, it reduces to the requirement
\be
\sum_i\left\llb \xi_i\otimes\xi_i,\tnu^2\right\rrb^{A}=2\tnu_B\sum_i\xi_i^A [\xi_i,\tnu]^B\,.
\ee
This is different to last condition of \eqref{eq: dual Sym conditions 0} (which as we have seen previously is given by \eqref{eq: Cas phi com}) because there is no symmetrization over the indices $\{A B\}$.
Thus for the dual metric ansatz we actually have a \emph{stronger} requirement for a symmetry to lift to the full spacetime. 

{As we have remarked in the main text, the Myers--Perry metrics with equal rotation parameters~\cite{Kunduri:2006qa} are in the form of \eqref{Generic dual}. We note moreover that, the Killing tensors of the Kerr--NUT--AdS black are almost of this form. See \cite{Kubiznak:2007ca} wherein, the Killing tensors are shown to be Killing tensors of the induced metric on the $t=$const. hypersurfaces, i.e. a co-dimension \emph{one} foliation.}

\subsection{Comparison with Taub--NUT metrics}
An example of a spacetime that takes also the form \eqref{Generic dual} is the Taub--NUT metric~\cite{Taub:1950ez,Newman:1963yy}
\begin{equation}
    ds^2=-f(r)(dt-2n\cos\theta d\phi)^2+\frac{dr^2}{f(r)}+(r^2+n^2)(d\theta^2+\sin^2\theta d\phi^2)\,,
\end{equation}
where $f(r)=(r^2+n^2-2mr)/(r^2+n^2)$. 
However, one can check that the corresponding vector
\begin{equation}
    \tnu=\frac{-2n\cos\theta}{(r^2+n^2)\sin^2\theta}\partial_\phi\,,
\end{equation}
is not a symmetry of the metric $\tgamma=\tgamma_{AB}dx^A dx^B=(r^2+n^2)(d\theta^2+\sin^2\!\theta d\phi^2)$. These are simply the Killing vectors of the 2-sphere
\be\label{KVSNUT}
    L_x=-\sin\phi\,\partial_\theta-\cos\phi\cot\theta\partial_\phi\,,
    \quad   L_y=\cos\phi\,\partial_\theta-\sin\phi\cot\theta\,\partial_\phi\,,\quad 
    L_z=\partial_\phi\,.
\ee
Note that, as mentioned in Subsection \ref{subsec: Taub--NUT}, the Taub--NUT metric actually admits 4 full Killing vectors---see, e.g. appendix A of \cite{Kalamakis:2020aaj} for details.
Moreover, the total angular momentum (our natural candidate for lifting) 
\begin{equation}
    C=L_x^2+L_y^2+L^2_z=\partial_\theta^2+\frac{1}{\sin^2\theta}\partial_\phi^2={(r^2+n^2)}\gamma^{-1}\,,
\end{equation}
does not commute with $\nu^A$, i.e.,
\be
\llb \nu, C\rrb_\SN\neq0\,.
\ee
Finally, since $C^{AB}=(r^2+n^2)\gamma^{AB}$ the third condition of \eqref{eq: dual Sym conditions 0} becomes simply that
\begin{equation}
    \partial_A(\nu^B\nu_B)=\frac{4n^2}{(r^2+n^2)}\partial_A(\cot^2\theta)\neq0\,,
\end{equation}
which is also explicitly violated.

Therefore, $\hat{C}$ is not one of the Killing tensors of the Taub--NUT spacetime. See, e.g. \cite{Jezierski:2006fw,Gibbons:1987sp} for a complete discussion of the symmetries of the Taub--NUT spacetime. We also mention that exactly the same calculations apply to the Page instanton~\cite{Page:1978vqj} (or any other such Euclideanization of the Taub--NUT), that is, the symmetries do \emph{not} lift from a base space according to the procedure discussed in this appendix.

\section{Schouten--Nijenhuis brackets}\label{App: SN brackets}

Here we collect various properties of the SN bracket defined over any manifold ${\cal N}$ which we employ in the main text. 
We use the early greek alphabet indices $\alpha,\beta,\gamma$ on $\cal N$ and note that the properties apply equally to quantities on $\cal M$ or $\cal S$. 
These properties of the SN bracket are due to the one-to-one correspondence with the Poisson bracket which satisfies the same kind of properties. 

Namely, consider the Poisson bracket on the space of functions on the cotangent bundle ${\cal T}^*{\cal N}$, defined in Darboux coordinates
\be
\{f,g\}\equiv\frac{\partial f}{\partial x^\alpha}\frac{\partial g}{\partial p_\alpha}-
\frac{\partial f}{\partial p_\alpha}\frac{\partial g}{\partial x^\alpha}\,.
\ee
Such a Poisson bracket is clearly antisymmetric and bilinear and satisfies the following properties:
\begin{align}
&\text{i) Jacobi:} \quad \{\{f,g\},h\}+\{\{h,f\},g\}+\{\{g,h\},f\}=0\,,
\\
& \text{ii) Leibniz:} \quad \{fg,h\}=\{f,h\}g +f\{g,h\}\,.
\end{align}
Thus, it defines a Leibniz Lie algebra.

We can then use the isomorsphism between the space of monomials on the contangent bundle and the space of symmetric tensors to define the SN bracket.%
\footnote{Unfortunately, in the mathematical community the SN bracket often refers to a similar construction but for anti-symmetric tensors. 
In the physics community this is almost never the case.}
That is,  for symmetric rank-$p$/rank-$q$ tensors $A^{\alpha_1\dots \alpha_p}$ and $B^{\alpha_1\dots \alpha_q}$ define their corresponding monomials of momentum on the cotangent bundle as $a=A^{\alpha_1\dots \alpha_p}p_{\alpha_1}\dots p_{\alpha_p}$ and $b=B^{\alpha_1\dots \alpha_q}p_{\alpha_1}\dots p_{\alpha_q}$. 
Then from their Poisson bracket
\be
c\equiv\{a,b\}=-[A,B]_\SN^{\alpha_1\dots \alpha_{p+q-1}}p_{\alpha_1}\dots p_{\alpha_{p+q-1}}\,,
\ee
we have the SN bracket naturally defined
\ba
[A,B]_{\mbox{\tiny SN}}^{\alpha_1\dots \alpha_{p+q-1}}&\equiv&
pA^{\gamma (\alpha_1\dots \alpha_{p-1}}\partial_\gamma B^{\alpha_p\dots \alpha_{p+q-1})} 
 -qB^{\gamma(\alpha_1\dots \alpha_{q-1}}\partial_\gamma A^{\alpha_q \dots \alpha_{q+p-1})}\,. 
\ea
For a scalar $A=\Phi$ it is clear that
\be
[A,\Phi]_\SN=pA^{\gamma \alpha_1\dots \alpha_{p-1}}\partial_\gamma\Phi\,.
\ee
Thus the SN bracket reduces to the Lie derivative when $A=\xi^\mu$ and in particular the vanishing of the SN bracket is equivalent to the Killing vector equation when also $B=g^{\mu\nu}$.
For our interests a $p$-symmetry of the manifold $\cal N$ is a $p$-symmetric tensor $K^{\mu_1,\dots,\mu_p}$ for which the SN bracket with the inverse metric, $G^{-1}$ on $\cal N$ vanishes:
\be
[K,G^{-1}]=0 \iff \nabla^{\cal N}_{(\alpha_1} K_{\alpha_2\dots \alpha_{p+1})}=0\,.
\ee
That is $K$ is a rank-$p$ Killing tensor. We also denote the vector space of rank-$p$ Killing tensors over $\cal M$ as ${\cal K}^p({\cal M})$.

Notice, that due to the isomorphism with the Poisson bracket a Killing tensor is equivalent to the quantity $k\equiv K^{\alpha_1\dots\alpha_p}p_{\alpha_1}\dots p_{\alpha_p}$ being a constant of motion for the geodesic Hamiltonian $g\equiv \frac{1}{2}G^{\alpha_1\alpha_2}p_{\alpha_1,\alpha_2}$
\be
\{k,g\}=0\,.
\ee

Now, the SN bracket defines a Lie algebra since it satisfies the Jacobi identity
\be\label{eq: SN Jacobi}
 [[A,B]_\SN,C]_\SN+[[C,A]_\SN,B]_\SN+[[B,C]_\SN,A]_\SN=0\,.
\ee
This implies that, a non-trivial SN bracket of any two symmetries $K_1$ and $K_2$ generates a new non-trivial symmetry, given by
$K_3 \equiv [K_1,K_2]_\SN\neq0$
\be
[K_3,G^{-1}]_\SN=[[K_1,K_2]_\SN,G^{-1}]_\SN=0\,.
\ee
Thus, we can keep generating new symmetries of the manifold by iterating SN bracket as in the main text.

Next, noting that the multiplication of two monomials of momenta corresponds to the symmetrized tensor product $\odot$ (e.g. for two vectors  $X\odot Y=\frac{1}{2}(X\otimes Y+Y\otimes X)$ of their corresponding tensors we have generically a Leibniz rule
\begin{align}\label{eq: Leibniz SN bracket proof}
\{ab,c\}&=\{a,c\}b+a\{b,c\}=-[A\odot B, C]_\SN^{\alpha_1\dots A_{p+q+r-1}} p_1\dots p_{\alpha_{p+q+r-1}}
\\
&=-([A,C]_\SN^{\alpha_1\dots \alpha_{p+q-1}}  B^{\alpha_{p+q}\dots \alpha_{p+q+r-1}}
        +A^{\alpha_1\dots \alpha_p}[B,C]_\SN^{\alpha_{p+1}\dots \alpha_{p+q+r-1}}
        )  p_1\dots p_{\alpha_{p+q+r-1}}\,,
\end{align}
so that
\be\label{eq: Leibniz SN bracket}
[A\odot B, C]_\SN=[A, C]_\SN\odot B+A\odot[B,C]_\SN\,.
\ee

Together, these properties mean that the SN bracket forms a graded Lie algebra (see e.g. \cite{MCLENAGHAN2004621,Yue2005}) on the space of Killing tensors of $\cal M$, 
\be
{\cal K}({\cal M}):=\bigoplus_{i=0} {\cal K}^i({\cal M})\,,
\ee
where ${\cal K}^{0}=\mathbb{R}$ and ${\cal K}^{1}$ is the space Killing vectors/isometries over $\cal M $.
Finally we remark, in manifolds of constant Riemann curvature, i.e. symmetric spaces, the representation of the algebra of Killing tensors is well studied, see e.g.~\cite{barbance1973tenseurs,DeLong1982,Takeuchi1983,Thompson1986,MCLENAGHAN2004621,Yue2005}. In particular, there is an isomorphism between a certain quotient of the universal covering algebra of the symmetry group, and the space of Killing tensors with the Lie bracket realized by the SN bracket (see Theorem 4.13 and corollary 4.14 of \cite{michel2017prolongation}).

\bibliography{workingbib}

\providecommand{\href}[2]{#2}\begingroup\raggedright\begin{thebibliography}{10}

\bibitem{Walker:1970un}
M.~Walker and R.~Penrose, \emph{{On quadratic first integrals of the geodesic
  equations for type $\{22\}$ spacetimes}},
  \href{https://doi.org/10.1007/BF01649445}{\emph{Commun. Math. Phys.}
  {\bfseries 18} (1970) 265}.

\bibitem{yano1952some}
K.~Yano, \emph{Some remarks on tensor fields and curvature},
  \href{https://doi.org/10.2307/1969782}{\emph{Annals of Mathematics} (1952)
  328}.

\bibitem{kerr1963gravitational}
R.~P. Kerr, \emph{Gravitational field of a spinning mass as an example of
  algebraically special metrics},
  \href{https://doi.org//10.1103/PhysRevLett.11.237}{\emph{Physical review
  letters} {\bfseries 11} (1963) 237}.

\bibitem{Myers:1986un}
R.~C. Myers and M.~J. Perry, \emph{{Black Holes in Higher Dimensional
  Space-Times}},
  \href{https://doi.org/10.1016/0003-4916(86)90186-7}{\emph{Annals Phys.}
  {\bfseries 172} (1986) 304}.

\bibitem{Gibbons:2004js}
G.~W. Gibbons, H.~L{\"u}, D.~N. Page and C.~N. Pope, \emph{{Rotating black
  holes in higher dimensions with a cosmological constant}},
  \href{https://doi.org/10.1103/PhysRevLett.93.171102}{\emph{Phys. Rev. Lett.}
  {\bfseries 93} (2004) 171102}
  [\href{https://arxiv.org/abs/hep-th/0409155}{{\ttfamily hep-th/0409155}}].

\bibitem{Gibbons:2004uw}
G.~W. Gibbons, H.~L{\"u}, D.~N. Page and C.~N. Pope, \emph{{The General
  Kerr--de Sitter metrics in all dimensions}},
  \href{https://doi.org/10.1016/j.geomphys.2004.05.001}{\emph{J. Geom. Phys.}
  {\bfseries 53} (2005) 49}
  [\href{https://arxiv.org/abs/hep-th/0404008}{{\ttfamily hep-th/0404008}}].

\bibitem{Carter:1968rr}
B.~Carter, \emph{{Global structure of the Kerr family of gravitational
  fields}}, \href{https://doi.org/10.1103/PhysRev.174.1559}{\emph{Phys. Rev.}
  {\bfseries 174} (1968) 1559}.

\bibitem{Carter:1977pq}
B.~Carter, \emph{{Killing Tensor Quantum Numbers and Conserved Currents in
  Curved Space}}, \href{https://doi.org/10.1103/PhysRevD.16.3395}{\emph{Phys.
  Rev. D} {\bfseries 16} (1977) 3395}.

\bibitem{Frolov:2006dqt}
V.~P. Frolov and D.~Kubiz\v{n}\'ak, \emph{{Hidden Symmetries of Higher
  Dimensional Rotating Black Holes}},
  \href{https://doi.org/10.1103/PhysRevLett.98.011101}{\emph{Phys. Rev. Lett.}
  {\bfseries 98} (2007) 011101}
  [\href{https://arxiv.org/abs/gr-qc/0605058}{{\ttfamily gr-qc/0605058}}].

\bibitem{Kubiznak:2006kt}
D.~Kubiz\v{n}\'ak and V.~P. Frolov, \emph{{Hidden Symmetry of Higher
  Dimensional Kerr-NUT-AdS Spacetimes}},
  \href{https://doi.org/10.1088/0264-9381/24/3/F01}{\emph{Class. Quant. Grav.}
  {\bfseries 24} (2007) F1}
  [\href{https://arxiv.org/abs/gr-qc/0610144}{{\ttfamily gr-qc/0610144}}].

\bibitem{Frolov:2017kze}
V.~Frolov, P.~Krtou{\v s} and D.~Kubiz\v{n}\'ak, \emph{{Black holes, hidden
  symmetries, and complete integrability}},
  \href{https://doi.org/10.1007/s41114-017-0009-9}{\emph{Living Rev. Rel.}
  {\bfseries 20} (2017) 6} [\href{https://arxiv.org/abs/1705.05482}{{\ttfamily
  1705.05482}}].

\bibitem{schouten1940uber}
J.~A. Schouten, \emph{Uber differentialkomitanten zweier kontravarianter
  grossen},  in \emph{Proc. Kon. Ned. Akad. Wet. Amsterdam}, vol.~43,
  pp.~449--452, 1940.

\bibitem{nijenhuis1955jacobi}
A.~Nijenhuis, \emph{Jacobi-type identities for bilinear differential
  concomitants of certain tensor fields}, {\emph{Indag. Math.} {\bfseries 17}
  (1955) 390}.

\bibitem{dolan1989significance}
P.~Dolan, A.~Kladouchou and C.~Card, \emph{On the significance of killing
  tensors}, \href{https://doi.org/10.1007/BF00760441}{\emph{General relativity
  and gravitation} {\bfseries 21} (1989) 427}.

\bibitem{barbance1973tenseurs}
C.~Barbance, \emph{Sur les tenseurs sym{\'e}triques}, {\emph{CR Acad. Sci.
  Paris S{\'e}r. AB} {\bfseries 276} (1973) A387}.

\bibitem{DeLong1982}
J.~Richard Peter~DeLong, \emph{Killing Tensors and the Hamilton--Jacobi
  Equation}, Ph.D. thesis, University of Minnesota, 1982.

\bibitem{Takeuchi1983}
M.~Takeuchi, \emph{{Killing tensor fields on spaces of constant curvature}},
  \href{https://doi.org/10.21099/tkbjm/1496159823}{\emph{Tsukuba Journal of
  Mathematics} {\bfseries 7} (1983) 233 }.

\bibitem{Thompson1986}
G.~Thompson, \emph{Killing tensors in spaces of constant curvature},
  \href{https://doi.org/10.1063/1.527288}{\emph{Journal of Mathematical
  Physics} {\bfseries 27} (1986) 2693}.

\bibitem{MCLENAGHAN2004621}
R.~G. McLenaghan, R.~Milson and R.~G. Smirnov, \emph{Killing tensors as
  irreducible representations of the general linear group},
  \href{https://doi.org/https://doi.org/10.1016/j.crma.2004.07.017}{\emph{Comptes
  Rendus Mathematique} {\bfseries 339} (2004) 621}.

\bibitem{Yue2005}
J.~Yue, \emph{The 1856 lemma of {C}ayley revisited. {I}. infinitesimal
  generators}, \href{https://doi.org/10.1063/1.1945747}{\emph{Journal of
  Mathematical Physics} {\bfseries 46} (2005) 073511}
  [\href{https://arxiv.org/abs/math/0407202}{{\ttfamily math/0407202}}].

\bibitem{michel2017prolongation}
J.-P. Michel, P.~Somberg and J.~{\v{S}}ilhan, \emph{Prolongation of symmetric
  {K}illing tensors and commuting symmetries of the {L}aplace operator},
  \href{https://doi.org/10.1216/RMJ-2017-47-2-587}{\emph{The Rocky Mountain
  Journal of Mathematics} {\bfseries 47} (2017) 587}
  [\href{https://arxiv.org/abs/1403.7226}{{\ttfamily 1403.7226}}].

\bibitem{Miller_2013}
W.~Miller, S.~Post and P.~Winternitz, \emph{Classical and quantum
  superintegrability with applications},
  \href{https://doi.org/10.1088/1751-8113/46/42/423001}{\emph{Journal of
  Physics A: Mathematical and Theoretical} {\bfseries 46} (2013) 423001}
  [\href{https://arxiv.org/abs/1309.2694}{{\ttfamily 1309.2694}}].

\bibitem{Gray:2021toe}
F.~Gray and D.~Kubiz\v{n}\'ak, \emph{{Slowly rotating black holes with exact
  Killing tensor symmetries}},
  \href{https://doi.org/10.1103/PhysRevD.105.064017}{\emph{Phys. Rev. D}
  {\bfseries 105} (2022) 064017}
  [\href{https://arxiv.org/abs/2110.14671}{{\ttfamily 2110.14671}}].

\bibitem{Gray:2021roq}
F.~Gray, R.~A. Hennigar, D.~Kubiz\v{n}\'ak, R.~B. Mann and M.~Srivastava,
  \emph{{Generalized Lense--Thirring metrics: higher-curvature corrections and
  solutions with matter}},
  \href{https://doi.org/10.1007/JHEP04(2022)070}{\emph{JHEP} {\bfseries 04}
  (2022) 070} [\href{https://arxiv.org/abs/2112.07649}{{\ttfamily
  2112.07649}}].

\bibitem{Gray:2024qys}
F.~Gray, C.~Keeler, D.~Kubiz\v{n}\'ak and V.~Martin, \emph{{Love symmetry in
  higher-dimensional rotating black hole spacetimes}},
  \href{https://doi.org/10.1007/JHEP03(2025)036}{\emph{JHEP} {\bfseries 03}
  (2025) 036} [\href{https://arxiv.org/abs/2409.05964}{{\ttfamily
  2409.05964}}].

\bibitem{Baines:2020unr}
J.~Baines, T.~Berry, A.~Simpson and M.~Visser,
  \emph{{Painlev\'e\textendash{}Gullstrand form of the
  Lense\textendash{}Thirring Spacetime}},
  \href{https://doi.org/10.3390/universe7040105}{\emph{Universe} {\bfseries 7}
  (2021) 105} [\href{https://arxiv.org/abs/2006.14258}{{\ttfamily
  2006.14258}}].

\bibitem{Baines:2021qaw}
J.~Baines, T.~Berry, A.~Simpson and M.~Visser, \emph{{Killing Tensor and Carter
  Constant for Painlev\'e\textendash{}Gullstrand Form of
  Lense\textendash{}Thirring Spacetime}},
  \href{https://doi.org/10.3390/universe7120473}{\emph{Universe} {\bfseries 7}
  (2021) 473} [\href{https://arxiv.org/abs/2110.01814}{{\ttfamily
  2110.01814}}].

\bibitem{Sadeghian:2022ihp}
S.~Sadeghian, \emph{{Killing tensors of a generalized Lense-Thirring
  spacetime}}, \href{https://doi.org/10.1103/PhysRevD.106.104028}{\emph{Phys.
  Rev. D} {\bfseries 106} (2022) 104028}
  [\href{https://arxiv.org/abs/2210.15431}{{\ttfamily 2210.15431}}].

\bibitem{Beig:1996ew}
R.~Beig and P.~T. Chrusciel, \emph{{Killing initial data}},
  \href{https://doi.org/10.1088/0264-9381/14/1A/007}{\emph{Class. Quant. Grav.}
  {\bfseries 14} (1997) A83}
  [\href{https://arxiv.org/abs/gr-qc/9604040}{{\ttfamily gr-qc/9604040}}].

\bibitem{Chrusciel:2013aya}
P.~T. Chru\'sciel and T.-T. Paetz, \emph{{KIDs like cones}},
  \href{https://doi.org/10.1088/0264-9381/30/23/235036}{\emph{Class. Quant.
  Grav.} {\bfseries 30} (2013) 235036}
  [\href{https://arxiv.org/abs/1305.7468}{{\ttfamily 1305.7468}}].

\bibitem{Kobialko:2021aqg}
K.~Kobialko, I.~Bogush and D.~Gal'tsov, \emph{{Killing tensors and photon
  surfaces in foliated spacetimes}},
  \href{https://doi.org/10.1103/PhysRevD.104.044009}{\emph{Phys. Rev. D}
  {\bfseries 104} (2021) 044009}
  [\href{https://arxiv.org/abs/2104.02167}{{\ttfamily 2104.02167}}].

\bibitem{Bogush:2021qnz}
I.~Bogush, K.~Kobialko and D.~Gal'tsov, \emph{{Killing tensors in foliated
  spacetimes and photon surfaces}},  in \emph{{16th Marcel Grossmann Meeting
  on~Recent Developments in Theoretical and Experimental General Relativity,
  Astrophysics and Relativistic Field Theories}}, 10, 2021,
  \href{https://arxiv.org/abs/2110.04608}{{\ttfamily 2110.04608}},
  \href{https://doi.org/10.1142/9789811269776_0317}{DOI}.

\bibitem{Kubiznak:2007ca}
D.~Kubiz\v{n}\'ak and V.~P. Frolov, \emph{{Stationary strings and branes in the
  higher-dimensional Kerr-NUT-(A)dS spacetimes}},
  \href{https://doi.org/10.1088/1126-6708/2008/02/007}{\emph{JHEP} {\bfseries
  02} (2008) 007} [\href{https://arxiv.org/abs/0711.2300}{{\ttfamily
  0711.2300}}].

\bibitem{Garfinkle:2010er}
D.~Garfinkle and E.~N. Glass, \emph{{Killing Tensors and Symmetries}},
  \href{https://doi.org/10.1088/0264-9381/27/9/095004}{\emph{Class. Quant.
  Grav.} {\bfseries 27} (2010) 095004}
  [\href{https://arxiv.org/abs/1003.0019}{{\ttfamily 1003.0019}}].

\bibitem{Garfinkle:2013cha}
D.~Garfinkle and E.~N. Glass, \emph{{Killing-Yano tensors in spaces admitting a
  hypersurface orthogonal Killing vector}},
  \href{https://doi.org/10.1063/1.4795122}{\emph{J. Math. Phys.} {\bfseries 54}
  (2013) 032501} [\href{https://arxiv.org/abs/1302.6207}{{\ttfamily
  1302.6207}}].

\bibitem{rajaratnam2014killing}
K.~Rajaratnam and R.~G. McLenaghan, \emph{Killing tensors, warped products and
  the orthogonal separation of the {H}amilton--{J}acobi equation},
  \href{https://doi.org/10.1063/1.4861707}{\emph{Journal of Mathematical
  Physics} {\bfseries 55} (2014) }
  [\href{https://arxiv.org/abs/1404.3161}{{\ttfamily 1404.3161}}].

\bibitem{Krtous:2015ona}
P.~Krtou{\v s}, D.~Kubiz\v{n}\'ak and I.~Kol\'a\v{r}, \emph{{Killing-Yano forms
  and Killing tensors on a warped space}},
  \href{https://doi.org/10.1103/PhysRevD.93.024057}{\emph{Phys. Rev. D}
  {\bfseries 93} (2016) 024057}
  [\href{https://arxiv.org/abs/1508.02642}{{\ttfamily 1508.02642}}].

\bibitem{Gibbons:2011hg}
G.~W. Gibbons, T.~Houri, D.~Kubiz\v{n}\'ak and C.~M. Warnick, \emph{{Some
  Spacetimes with Higher Rank Killing-Stackel Tensors}},
  \href{https://doi.org/10.1016/j.physletb.2011.04.047}{\emph{Phys. Lett. B}
  {\bfseries 700} (2011) 68} [\href{https://arxiv.org/abs/1103.5366}{{\ttfamily
  1103.5366}}].

\bibitem{Clement:2002mb}
G.~Clement, D.~Gal'tsov and C.~Leygnac, \emph{{Linear dilaton black holes}},
  \href{https://doi.org/10.1103/PhysRevD.67.024012}{\emph{Phys. Rev. D}
  {\bfseries 67} (2003) 024012}
  [\href{https://arxiv.org/abs/hep-th/0208225}{{\ttfamily hep-th/0208225}}].

\bibitem{Vasudevan:2004ca}
M.~Vasudevan, K.~A. Stevens and D.~N. Page, \emph{{Separability of the
  Hamilton-Jacobi and Klein-Gordon equations in Kerr--de Sitter metrics}},
  \href{https://doi.org/10.1088/0264-9381/22/2/007}{\emph{Class. Quant. Grav.}
  {\bfseries 22} (2005) 339}
  [\href{https://arxiv.org/abs/gr-qc/0405125}{{\ttfamily gr-qc/0405125}}].

\bibitem{barut1986theory}
A.~Barut and R.~Rączka, \emph{Theory of Group Representations and
  Applications}. World Scientific, 1986.

\bibitem{robson1988noncommutative}
J.~Robson, \emph{Noncommutative Noetherian Rings}, Wiley series in pure and
  applied mathematics. Wiley, 1988.

\bibitem{iachello2006lie}
F.~Iachello, \emph{Lie Algebras and Applications}, Lecture Notes in Physics.
  Springer Berlin Heidelberg, 2006.

\bibitem{Houri:2017tlk}
T.~Houri, K.~Tomoda and Y.~Yasui, \emph{{On integrability of the Killing
  equation}}, \href{https://doi.org/10.1088/1361-6382/aaa4e7}{\emph{Class.
  Quant. Grav.} {\bfseries 35} (2018) 075014}
  [\href{https://arxiv.org/abs/1704.02074}{{\ttfamily 1704.02074}}].

\bibitem{Chrusciel:1993hx}
P.~T. Chrusciel, M.~A.~H. MacCallum and D.~B. Singleton, \emph{{Gravitational
  waves in general relativity: 14. Bondi expansions and the polyhomogeneity of
  Scri}}, \href{https://doi.org/10.1098/rsta.1995.0004}{\emph{Philosophical
  Transactions of the Royal Society of London. Series A: Physical and
  Engineering Sciences} {\bfseries 350} (1995) 113}
  [\href{https://arxiv.org/abs/gr-qc/9305021}{{\ttfamily gr-qc/9305021}}].

\bibitem{Bazanski:1990qd}
S.~L. Bazanski and P.~Zyla, \emph{{A Gauss type law for gravity with a
  cosmological constant}}, \href{https://doi.org/10.1007/BF00756146}{\emph{Gen.
  Rel. Grav.} {\bfseries 22} (1990) 379}.

\bibitem{Kastor:2008xb}
D.~Kastor, \emph{{Komar Integrals in Higher (and Lower) Derivative Gravity}},
  \href{https://doi.org/10.1088/0264-9381/25/17/175007}{\emph{Class. Quant.
  Grav.} {\bfseries 25} (2008) 175007}
  [\href{https://arxiv.org/abs/0804.1832}{{\ttfamily 0804.1832}}].

\bibitem{Myers:2011yc}
R.~C. Myers, \emph{{Myers\textendash{}Perry black holes}},  in \emph{{Black
  holes in higher dimensions}} (G.~T. Horowitz, ed.), pp.~101--133.
\newblock Cambridge University Press, 2012.
\newblock \href{https://arxiv.org/abs/1111.1903}{{\ttfamily 1111.1903}}.

\bibitem{Galajinsky:2013mla}
A.~Galajinsky, A.~Nersessian and A.~Saghatelian, \emph{{Superintegrable models
  related to near horizon extremal Myers-Perry black hole in arbitrary
  dimension}}, \href{https://doi.org/10.1007/JHEP06(2013)002}{\emph{JHEP}
  {\bfseries 06} (2013) 002} [\href{https://arxiv.org/abs/1303.4901}{{\ttfamily
  1303.4901}}].

\bibitem{Sen:1992ua}
A.~Sen, \emph{{Rotating charged black hole solution in heterotic string
  theory}}, \href{https://doi.org/10.1103/PhysRevLett.69.1006}{\emph{Phys. Rev.
  Lett.} {\bfseries 69} (1992) 1006}
  [\href{https://arxiv.org/abs/hep-th/9204046}{{\ttfamily hep-th/9204046}}].

\bibitem{Galtsov:1994pd}
D.~V. Galtsov and O.~V. Kechkin, \emph{{Ehlers-Harrison type transformations in
  dilaton - axion gravity}},
  \href{https://doi.org/10.1103/PhysRevD.50.7394}{\emph{Phys. Rev. D}
  {\bfseries 50} (1994) 7394}
  [\href{https://arxiv.org/abs/hep-th/9407155}{{\ttfamily hep-th/9407155}}].

\bibitem{Garcia:1995qz}
A.~Garcia, D.~Galtsov and O.~Kechkin, \emph{{Class of stationary axisymmetric
  solutions of the Einstein-Maxwell dilaton - axion field equations}},
  \href{https://doi.org/10.1103/PhysRevLett.74.1276}{\emph{Phys. Rev. Lett.}
  {\bfseries 74} (1995) 1276}.

\bibitem{Galtsov:2025nia}
D.~Gal'tsov and R.~Karsanov, \emph{{Gauged supergravities: Solutions with a
  Killing tensor}},
  \href{https://doi.org/10.1103/PhysRevD.111.104011}{\emph{Phys. Rev. D}
  {\bfseries 111} (2025) 104011}
  [\href{https://arxiv.org/abs/2503.06589}{{\ttfamily 2503.06589}}].

\bibitem{Chrusciel:2025zfv}
P.~T. Chru\'sciel, W.~Cong and F.~Gray, \emph{{Kerr-AdS type higher dimensional
  black holes with non-spherical cross-sections of horizons}},
  \href{https://arxiv.org/abs/2501.05543}{{\ttfamily 2501.05543}}.

\bibitem{Kunduri:2006qa}
H.~K. Kunduri, J.~Lucietti and H.~S. Reall, \emph{{Gravitational perturbations
  of higher dimensional rotating black holes: Tensor perturbations}},
  \href{https://doi.org/10.1103/PhysRevD.74.084021}{\emph{Phys. Rev. D}
  {\bfseries 74} (2006) 084021}
  [\href{https://arxiv.org/abs/hep-th/0606076}{{\ttfamily hep-th/0606076}}].

\bibitem{Vasudevan:2004mr}
M.~Vasudevan, K.~A. Stevens and D.~N. Page, \emph{{Particle motion and scalar
  field propagation in Myers-Perry black hole spacetimes in all dimensions}},
  \href{https://doi.org/10.1088/0264-9381/22/7/017}{\emph{Class. Quant. Grav.}
  {\bfseries 22} (2005) 1469}
  [\href{https://arxiv.org/abs/gr-qc/0407030}{{\ttfamily gr-qc/0407030}}].

\bibitem{He:2024tab}
S.~He and Y.~Li, \emph{{Spacetimes with prescribed Killing tensor symmetries}},
  \href{https://doi.org/10.1103/PhysRevD.110.084076}{\emph{Phys. Rev. D}
  {\bfseries 110} (2024) 084076}
  [\href{https://arxiv.org/abs/2407.11178}{{\ttfamily 2407.11178}}].

\bibitem{Frausto:2024egp}
G.~Frausto, I.~Kol\'a\v{r}, T.~M\'alek and C.~Torre, \emph{{Symmetry reduction
  of gravitational Lagrangians}},
  \href{https://doi.org/10.1103/PhysRevD.111.064062}{\emph{Phys. Rev. D}
  {\bfseries 111} (2025) 064062}
  [\href{https://arxiv.org/abs/2410.11036}{{\ttfamily 2410.11036}}].

\bibitem{Taub:1950ez}
A.~H. Taub, \emph{{Empty space-times admitting a three parameter group of
  motions}}, \href{https://doi.org/10.2307/1969567}{\emph{Annals Math.}
  {\bfseries 53} (1951) 472}.

\bibitem{Newman:1963yy}
E.~Newman, L.~Tamburino and T.~Unti, \emph{{Empty space generalization of the
  Schwarzschild metric}}, \href{https://doi.org/10.1063/1.1704018}{\emph{J.
  Math. Phys.} {\bfseries 4} (1963) 915}.

\bibitem{Gibbons:1987sp}
G.~W. Gibbons and P.~J. Ruback, \emph{{The Hidden Symmetries of Multicenter
  Metrics}}, \href{https://doi.org/10.1007/BF01466773}{\emph{Commun. Math.
  Phys.} {\bfseries 115} (1988) 267}.

\bibitem{Jezierski:2006fw}
J.~Jezierski and M.~Lukasik, \emph{{Conformal Yano-Killing tensors for the
  Taub--NUT metric}},
  \href{https://doi.org/10.1088/0264-9381/24/5/015}{\emph{Class. Quant. Grav.}
  {\bfseries 24} (2007) 1331}
  [\href{https://arxiv.org/abs/gr-qc/0610090}{{\ttfamily gr-qc/0610090}}].

\bibitem{Kalamakis:2020aaj}
G.~Kalamakis, R.~G. Leigh and A.~C. Petkou, \emph{{Aspects of holography of
  Taub--NUT-AdS$_4$ spacetimes}},
  \href{https://doi.org/10.1103/PhysRevD.103.126012}{\emph{Phys. Rev. D}
  {\bfseries 103} (2021) 126012}
  [\href{https://arxiv.org/abs/2009.08022}{{\ttfamily 2009.08022}}].

\bibitem{Awad:2000gg}
A.~Awad and A.~Chamblin, \emph{{A Bestiary of higher dimensional Taub--NUT-AdS
  space-times}}, \href{https://doi.org/10.1088/0264-9381/19/8/301}{\emph{Class.
  Quant. Grav.} {\bfseries 19} (2002) 2051}
  [\href{https://arxiv.org/abs/hep-th/0012240}{{\ttfamily hep-th/0012240}}].

\bibitem{Bueno:2018uoy}
P.~Bueno, P.~A. Cano, R.~A. Hennigar and R.~B. Mann, \emph{{NUTs and bolts
  beyond Lovelock}}, \href{https://doi.org/10.1007/JHEP10(2018)095}{\emph{JHEP}
  {\bfseries 10} (2018) 095}
  [\href{https://arxiv.org/abs/1808.01671}{{\ttfamily 1808.01671}}].

\bibitem{Cong:2024xam}
W.~Cong, P.~T. Chru\'sciel and F.~Gray, \emph{{Characteristic Gluing with
  $\Lambda$: II. Linearised equations in higher dimensions}},
  \href{https://arxiv.org/abs/2401.04442}{{\ttfamily 2401.04442}}.

\bibitem{Moncrief:1983xua}
V.~Moncrief and J.~Isenberg, \emph{{Symmetries of cosmological Cauchy
  horizons}}, \href{https://doi.org/10.1007/BF01214662}{\emph{Commun. Math.
  Phys.} {\bfseries 89} (1983) 387}.

\bibitem{Bondi:1962px}
H.~Bondi, M.~G.~J. van~der Burg and A.~W.~K. Metzner, \emph{{Gravitational
  waves in general relativity. 7. Waves from axisymmetric isolated systems}},
  \href{https://doi.org/10.1098/rspa.1962.0161}{\emph{Proc. Roy. Soc. Lond. A}
  {\bfseries 269} (1962) 21}.

\bibitem{Sachs:1962zza}
R.~K. Sachs, \emph{{Gravitational waves in general relativity. 8. Waves in
  asymptotically flat space-times}},
  \href{https://doi.org/10.1098/rspa.1962.0206}{\emph{Proc. Roy. Soc. Lond. A}
  {\bfseries 270} (1962) 103}.

\bibitem{Strominger:2017zoo}
A.~Strominger, \emph{{Lectures on the Infrared Structure of Gravity and Gauge
  Theory}}, {\emph{Princeton University Press} (2018) }
  [\href{https://arxiv.org/abs/1703.05448}{{\ttfamily 1703.05448}}].

\bibitem{Campoleoni:2021blr}
A.~Campoleoni and S.~Pekar, \emph{{Carrollian and Galilean conformal
  higher-spin algebras in any dimensions}},
  \href{https://doi.org/10.1007/JHEP02(2022)150}{\emph{JHEP} {\bfseries 02}
  (2022) 150} [\href{https://arxiv.org/abs/2110.07794}{{\ttfamily
  2110.07794}}].

\bibitem{Barnich:2005bn}
G.~Barnich, N.~Bouatta and M.~Grigoriev, \emph{{Surface charges and dynamical
  Killing tensors for higher spin gauge fields in constant curvature spaces}},
  \href{https://doi.org/10.1088/1126-6708/2005/10/010}{\emph{JHEP} {\bfseries
  10} (2005) 010} [\href{https://arxiv.org/abs/hep-th/0507138}{{\ttfamily
  hep-th/0507138}}].

\bibitem{Odak:2023jgs}
G.~Odak, \emph{{Polarizations in phase space: Boundary conditions and
  gravitational charges}}, Ph.D. thesis, Aix-Marseille Universit\'e, 9, 2023.

\bibitem{Page:1978vqj}
D.~N. Page, \emph{{A compact rotating gravitational instanton}},
  \href{https://doi.org/10.1016/0370-2693(78)90231-9}{\emph{Phys. Lett. B}
  {\bfseries 79} (1978) 235}.

\end{thebibliography}\endgroup
\bibliographystyle{JHEP}

\end{document}